\documentclass[aip,jcp,reprint,noshowkeys]{revtex4-1}
\usepackage{graphicx,float,dcolumn,bm,xcolor,microtype,multirow,amscd,amsmath,amssymb,amsfonts,physics,mhchem,upgreek,pifont,wrapfig,enumitem,siunitx}
\usepackage{libertine}
\usepackage[normalem]{ulem}

\usepackage{hyperref}
\hypersetup{
    colorlinks=true,
    linkcolor=blue,
    filecolor=blue,      
    urlcolor=blue,	
    citecolor=blue
}

\DeclareSIUnit{\angstrom}{\mbox{\normalfont\AA}}

\UndeclareTextCommand{\l}{OT1}
\DeclareTextSymbolDefault{\l}{T1}

\makeatletter
\DeclareRobustCommand{\AA}{%
	\leavevmode
	\vbox{\ialign{##\cr
			\hidewidth\char'27 \hidewidth\cr
			\noalign{\nointerlineskip\kern-1.4ex}
			A\cr
	}}%
}
\makeatother

\usepackage{natbib}
\AtBeginDocument{\nocite{achemso-control}}
  
\newcolumntype{d}[1]{D{.}{.}{#1}}
\newcommand{\head}[1]{\multicolumn{1}{c}{#1}}
\newcommand\Tstrut{\rule{0pt}{2.5ex}}         % = `top' strut
\definecolor{hughgreen}{HTML}{009900}

\newcommand{\up}[1]{\,^{#1}}

\newcommand{\hH}{\hat{H}}
\newcommand{\Href}{\hH_{0}}
\newcommand{\Hpert}{\hat{V}}
\newcommand{\Heff}{\hH_{\text{eff}}}

\newcommand{\hFg}{\hat{F}_{\text{G}}}
\newcommand{\Fg}{F_{\text{G}}}

\newcommand{\cP}{\mathcal{P}}
\newcommand{\cQ}{\mathcal{Q}}
\newcommand{\cI}{\mathcal{I}}

\newcommand{\cL}{\mathcal{L}}
\newcommand{\cX}{\mathcal{X}}
\newcommand{\cY}{\mathcal{Y}}
\newcommand{\cW}{\mathcal{W}}

\newcommand{\pModel}{\cP}
\newcommand{\pExt}{\cQ}
\newcommand{\Pg}{P_{\text{G}}}

\newcommand{\brkt}[2]{\langle #1 | #2 \rangle}
\newcommand{\Wfn}{\Psi}
\newcommand{\wAO}{\chi}	% Atomic orbitals
\newcommand{\wMO}{\phi}
\newcommand{\wSCF}{\Phi}	% Multiple SCF determinants (non-orthogonal)
\newcommand{\wNOCI}{\Wfn^{\text{NOCI}}}	% ''True'' many-particle wave functions
\newcommand{\sig}{\upsigma}
\newcommand{\sigg}{\sig_{\text{g}}}
\newcommand{\sigu}{\sig_{\text{u}}}

\newcommand{\xNOCI}{c}		% Expansion of NOCI in SCF solution basis
\newcommand{\xMO}{C}        % Molecular orbital expansion coefficients
\newcommand{\xFO}{a}
\newcommand{\Nref}{n_{\text{ref}}}
\newcommand{\Ndet}{n_{\text{det}}}

\newcommand{\Ne}{N}

\newcommand{\Nbas}{n_\text{bas}}

\newcommand{\I}{\mathrm{i}}
\newcommand{\Eh}{\text{E}_{\text{h}}}
\newcommand{\mEh}{\text{m}\Eh}
	
\newcommand{\br}{\bm{r}}
\newcommand{\ba}{\bm{a}}
\newcommand{\bF}{\bm{F}}
\newcommand{\bI}{\bm{I}}
\newcommand{\bM}{\bm{M}}
\newcommand{\bP}{\bm{P}}
\newcommand{\bQ}{\bm{Q}}
\newcommand{\bV}{\bm{V}}
\newcommand{\bX}{\bm{X}}
\newcommand{\bY}{\bm{Y}}
\newcommand{\bDelta}{\bm{\Delta}}
\newcommand{\invbDelta}{\bDelta^{\hspace{-2pt}-1}}
\newcommand{\bPg}{\bP_{\text{G}}}

\newcommand{\bsM}{\boldsymbol{\mathcal{M}}}

\newcommand{\Or}{\mathcal{O}}

\newcommand{\oneE}{h}
\newcommand{\twoEas}[4]{\langle #1 #2 || #3 #4 \rangle}

\renewcommand{\bra}[1]{\langle #1 |}
\renewcommand{\ket}[1]{|#1\rangle}
\renewcommand{\braket}[3]{\langle #1 | #2 | #3 \rangle}

\newcommand{\qchem}{\textsc{Q-Chem~5.2}}
\newcommand{\libnoci}{\textsc{LIBNOCI}}
\newcommand{\openmolcas}{\textsc{OpenMolcas}}
\newcommand{\orca}{\textsc{ORCA}}
\newcommand{\mrcc}{\textsc{MRCC}}
\newcommand{\qp}{\textsc{Quantum~Package~2.0}}

\newcommand{\etal}{\textit{et al}}
\newcommand{\ie}{i.e.}
\newcommand{\ansatz}{\textit{ansatz}}

\usepackage{stackengine}
\stackMath
\newcommand\tsup[2][2]{%
	\def\useanchorwidth{T}%
	\ifnum#1>1%
	\stackon[-1.2ex]{\tsup[\numexpr#1-1\relax]{#2}}{\mathchar"307E}%
	\else%
	\stackon[-1ex]{#2}{\mathchar"307E}%
	\fi%
}
\renewcommand{\tilde}[1]{\tsup[1]{#1}}
\newcommand{\ttilde}[1]{\tsup[2]{#1}}
\newcommand{\UCAM}{Department of Chemistry, University of Cambridge, Lensfield Road, Cambridge, CB2 1EW, U.K.}

\begin{document}

\title{Reaching Full Correlation through Nonorthogonal Configuration Interaction:\\ A Second-Order Perturbative Approach}
%\title{Nonorthogonal Second-Order Multireference Perturbation Theory}
%\title{\hugh{Second-Order Perturbation Thoery for Nonorthogonal Configuration Interaction}}
\author{Hugh~G.~A.~Burton}
\email{hb407@cam.ac.uk}
\affiliation{\UCAM}
\author{Alex~J.~W.~Thom}
\email{ajwt3@cam.ac.uk}
\affiliation{\UCAM}
\date{\today}

\begin{abstract}
\begin{wrapfigure}[11]{r}{0.4\textwidth}
    \flushleft
    \vspace{-0.4cm}
    \hspace{-1.5cm}
    \fbox{\includegraphics[width=0.39\textwidth]{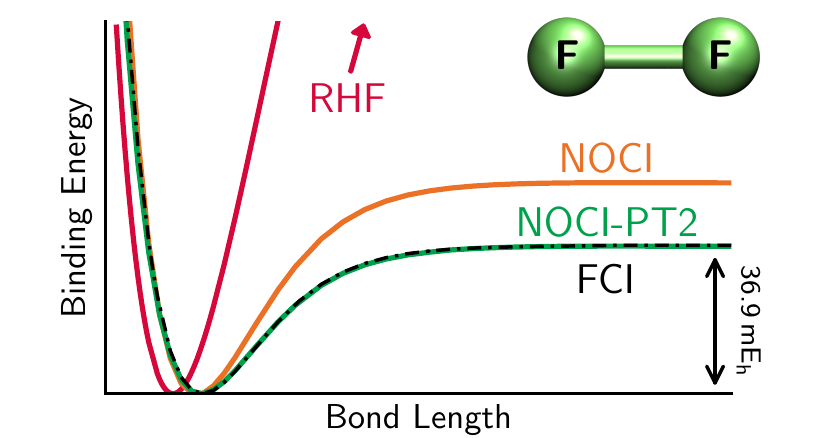}}
\end{wrapfigure}
Nonorthogonal multireference methods can predict statically correlated adiabatic energies while providing chemical insight through the combination of diabatic reference states. 
However, reaching quantitative accuracy using nonorthogonal multireference expansions remains a significant challenge.
In this work, we present the first rigorous perturbative correction to nonorthogonal configuration interaction, allowing the remaining dynamic correlation to be reliably computed.
Our second-order ``NOCI-PT2'' theory exploits a zeroth-order generalised Fock Hamiltonian and 
builds the first-order interacting space using single and double excitations from each reference determinant.
This approach therefore
defines the rigorous nonorthogonal extension to conventional multireference perturbation theory.
We find that NOCI-PT2 can quantitatively predict multireference potential energy surfaces and provides state-specific ground and excited states for adiabatic avoided crossings.
Furthermore, we introduce an explicit imaginary-shift formalism requiring shift values that are an order of magnitude smaller than those used in conventional multireference perturbation theory.
%Ultimately, these developments establish the nonorthogonal multireference framework as a reliable approach for predicting quantitative electronic energies.
\end{abstract}
\maketitle

\clearpage

%%%%%%%%%%%%%%%%%%%%%%%%%%%%%%%%%%%%%%%%%%%%%%%%%%%%%%
\section{Introduction}
%%%%%%%%%%%%%%%%%%%%%%%%%%%%%%%%%%%%%%%%%%%%%%%%%%%%%%
% NONORTHOGONAL METHODS
Nonorthogonal methods have a long history in quantum chemistry.
Starting from the Heitler--London wave function,\cite{Heitler1927} nonorthogonal orbitals have been exploited in valence bond methods,\cite{ShaikBook}  localised  population analysis,\cite{Weinhold1988} and the construction of compact multireference wave functions.\cite{Olsen2015,Kahler2017,Pathak2018}
Alternatively, the interaction of nonorthogonal many-electron basis functions can describe intermolecular forces,\cite{StoneBook} magnetic interactions,\cite{Broer2003} and electron transfer processes.\cite{Wibowo2017,Jensen2018,Kathir2020}
In many of these cases, the diabatic nature of nonorthogonal functions provides chemically intuitive results that are lost using strictly orthogonal techniques.

% NOCI PRELIM
One nonorthogonal approach that has recently received particular attention is the combination of multiple Hartree--Fock (HF) solutions to construct statically correlated multireference wave functions.\cite{Thom2009,Sundstrom2014}
As a nonlinear method, the HF equations can yield several solutions that correspond to local minima, maxima, and saddle points of the HF energy, each with a bespoke set of molecular orbitals.\cite{Kowalski1998,Thom2008}
The development of computational methods such as SCF metadynamics\cite{Thom2008} and the Maximum Overlap Method (MOM)\cite{Gilbert2008,Barca2018} has revived interest in these multiple stationary solutions.
For example, HF states have been proposed as mean-field excited states,\cite{Gilbert2008,Besley2009,Barca2014,Barca2018} and are found to behave as diabatic states along reaction pathways.\cite{Thom2009,Jensen2018}
However, HF wave functions are not eigenfunctions of the true molecular Hamiltonian and can break the spin $\hat{\mathcal{S}}^2$ and $\hat{\mathcal{S}}_z$ symmetries, or spatial point group symmetries.\cite{StuberBook,Jimenez-Hoyos2011}
The existence of multiple symmetry-broken HF solutions is a well-known indicator of static correlation in the exact wave function, with the emergence of the unrestricted HF (UHF) state at the  Coulson--Fischer point in \ce{H2} providing the typical example.\cite{Coulson1949}

% NOCI DESCRIPTION
The diabatic nature of HF states, and their links to static correlation, makes multiple HF solutions the ideal building blocks for multiconfigurational wave functions.
Since each HF solution is built from an individually optimised set of orbitals, multiple HF states are not mutually orthogonal and their linear combination takes the form of a nonorthogonal configuration interaction (NOCI).\cite{Thom2009}
The resulting NOCI wave functions can capture static correlation with a similar accuracy to the conventional orthogonal alternative:\cite{Mayhall2014,Sundstrom2014,Burton2019c} the complete active space self-consistent field (CASSCF) approach.\cite{Roos1980}
These NOCI expansions have also been found to produce adiabatic conical intersections and avoided crossings, and can provide size-consistent energies for bond dissociation.\cite{Thom2009}
Furthermore, including all symmetry-related HF solutions in the NOCI basis leads to the restoration of broken symmetries\cite{Huynh2019} in a similar manner to projected HF techniques.\cite{Smeyers1973,Smeyers1974,Jimenez-Hoyos2012} 

% CHALLENGING OPTIMISATION
The nonorthogonal structure of the NOCI expansion leads to a number of computational and chemical advantages over the CASSCF approach.
Firstly, the individual HF wave functions in the NOCI basis are individually optimised at the mean-field level prior to being combined in the NOCI expansion.\cite{Thom2009}
In contrast, the configurations in the CASSCF expansion are built from a common set of orbitals, and the wave function must be simultaneously optimised with respect to the orbital and configuration interaction (CI) coefficients.\cite{HelgakerBook}
CASSCF can therefore contain redundant parameters or convergence issues that require second-order optimisation schemes,\cite{Yeager1980,Werner1980,Olsen1982,Werner1984,Werner1985} 
while NOCI avoids any challenging optimisation and retains the orbital relaxation of different configurations.

% NOCI AVOIDS ACTIVE SPACE DEFINITION
Secondly, the NOCI wave function is built from many-electron  stationary states of the HF energy,\cite{Thom2009} while the CASSCF configurations are constructed from a common active set of single-electron orbitals.\cite{Roos1980}
As a result, the NOCI expansion functions vary smoothly along a potential energy surface, while the CASSCF active space must be defined using ``chemical intuition'' and can change discontinuously as the molecular structure evolves.\cite{Keller2015}
Furthermore, the CASSCF active space is sensitive to occupied-occupied orbital rotations in the reference determinant,\cite{HelgakerBook} whereas each NOCI basis function is well defined by the HF energy landscape.
Although real HF states can also disappear along a potential energy surface,\cite{Coulson1949,Thom2009} the introduction of the holomorphic HF approach has provided complex-analytic extensions of states that define a continuous basis for NOCI across all molecular structures.\cite{Hiscock2014, Burton2016, Burton2018, Burton2019c}

% EXCITED STATES
Finally, including individually optimised excited-state determinants in the NOCI basis allows orbital relaxation to be captured in state-specific valence\cite{Sundstrom2014} or core excitations,\cite{Oosterbaan2018,Oosterbaan2019} charge transfer processes,\cite{Jensen2018} and transition-metal complexes.\cite{Mayhall2014,Huynh2019}
Although state-specific excited states can be targeted in the CASSCF framework, these calculations are susceptible to ``root-flipping'' in the CI component of a CASSCF iteration that leads to variational collapse onto a lower-energy state.\cite{Tran2019}
Instead, a state-averaged CASSCF formalism is generally required, where a weighted average of the ground- and excited-state energies is optimised with a common set of active-space orbitals.\cite{Werner1981}
However, state-averaging leads to ground- and excited-state CASSCF wave functions that are not true stationary points of the CASSCF energy\cite{Tran2019} and relies on the arbitrary definition of optimisation weights for each target state.
Furthermore, the diabatic nature of multiple HF states enables the chemical interpretation of adiabatic NOCI states in the vicinity of conical intersections and avoided crossings,\cite{Thom2009,Jensen2018} while recovering a diabatic representation in state-averaged CASSCF is challenging.\cite{Domcke1993,Domcke1994,Fulscher2002}

% RESONATING HARTREE-FOCK
Orbital relaxation effects can be further enhanced by simultaneously optimising the nonorthogonal CI coefficients and the individual Slater determinant orbitals in the Resonating HF (Res-HF) approach.\cite{Fukutome1988,Ikawa1993}
Res-HF theory forms the nonorthogonal analogue of multiconfigurational self-consistent-field (MCSCF)  methods and can provide equivalent accuracy to a full valence CASSCF wave function.\cite{Tomita1996}
However, simultaneously optimising NOCI and orbital coefficients in the Res-HF wave function leads to similar challenges as the CASSCF wave function, while using nonorthogonal  determinants adds the possibility of two NOCI basis states coalescing during the optimisation procedure.\cite{Ikawa1993}
Despite these challenges, the Res-HF approach has seen a recent resurgence in the context of excited-state calculations, where orbital optimisation can sometimes provide significantly more accurate excitation energies.\cite{Nite2019}

% CHALLENGES
The applications of NOCI across chemistry have generally been limited by the lack of any significant dynamic electron correlation in the NOCI energy.\cite{Mayhall2014,Sundstrom2014,Burton2019c}
Like CASSCF, additional dynamic correlation can in principle be captured by using an increasingly large NOCI expansion basis, but these longer expansions remove much of the benefit provided by the compact NOCI wave function.
In the CASSCF framework, the most popular way to add dynamic correlation is to introduce a second-order perturbation theory (PT2) correction through the ``diagonalise-then-perturb'' CASPT2 approach.\cite{Andersson1990,Andersson1992}
However, the construction of a similar diagonalise-then-perturb correction for NOCI  has so far proved elusive.
%In this paper, we present the first derivation of a rigorous NOCI-based second-order perturbation theory that allows quantitative accuracy to be reached through the NOCI framework.

% PREVIOUS DYNAMIC CORRELATION
The primary obstacle to deriving dynamically-correlated post-NOCI techniques is the lack of a common set of molecular orbitals for the NOCI reference determinants.
While multireference CI and coupled-cluster (CC) extensions to NOCI were reported in Ref.~\onlinecite{TenNo1997}, subsequent developments have focussed on adding dynamic correlation to the NOCI expansion in an approximate manner.
For example, in the NOCI-MP2 approach, each reference determinant is individually perturbed using second-order M\o{}ller--Plesset perturbation theory\cite{Moller1934} (MP2) before the NOCI eigenvalue problem is solved, leading to a ``perturb-then-diagonalise'' approach.\cite{Yost2013,Yost2016,Yost2019}
However, the original NOCI-MP2 algorithm was found to contain size-consistency errors that required \textit{ad hoc} alterations to the working equations.\cite{Yost2016}
Alternatively, Nite and Jim\'{e}nez--Hoyos have recently reported an extension to NOCI that interacts CISD wave functions built from each reference determinant.\cite{Nite2019b}
Like NOCI-MP2, the resulting ``NOCISD'' approach introduces dynamic correlation before the eigenvalue problem is solved, and the NOCI coefficients are identified ``in the presence of'' the dynamic correlation.
However, both of these perturb-then-diagonalise methods rely on the existence of well-behaved MP2 or CISD expansions for the NOCI basis states, and will break down when the constituent MP2 or CISD wave functions fail.

% OUR PLAN AND SUCCESS
In this paper, we present the first derivation of a rigorous nonorthogonal second-order perturbation theory that allows quantitative accuracy to be reliably reached through the NOCI framework.
The resulting theory --- which we call NOCI-PT2 --- provides two key advantages over previous dynamically-correlated NOCI methods.
Firstly, by directly perturbing the NOCI wave function, NOCI-PT2 can be applied without relying on any properties of the NOCI basis states, such as the quality of their MP2 or CISD wave functions.
Secondly, NOCI-PT2 is essentially equivalent to CASPT2 when the NOCI basis contains a suitable set of orthogonal configurations, and reduces to MP2 when only one reference determinant is considered.
Our approach therefore provides the rigorous nonorthogonal extension to multireference perturbation theories such as CASPT2 and allows quantitative accuracy to be reached while exploiting the advantages of the NOCI framework.

% CONTENTS SUMMARY
Before deriving NOCI-PT2 theory, we first outline the prerequisite details of NOCI, holomorphic HF, and second-order perturbation theory in Section~\ref{sec:BackgroundTheory}.
We then develop the rigorous NOCI-PT2 correction in Section~\ref{sec:TheoryDevelopment} and discuss its computational implementation in Section~\ref{sec:ComputationalDetails}.
In Section~\ref{sec:Results}, we assess the performance of NOCI-PT2 by considering a series of typically challenging molecules with strong static and dynamic correlation.
Finally, we summarise our key findings and highlight future avenues of research in Section~\ref{sec:ConcludingRemarks}.

% CHOICE OF NOCI BASIS
%\hugh{\bf HB: I think this might be a slow distraction?}
%While the NOCI expansion can be built from \emph{any} nonorthogonal basis, there is no obvious `black-box' method to identify a suitable set of basis functions and various approaches have been proposed.
%The simplest approaches involve taking orthogonal excitations from a reference determinant and individually optimising each excitation at the HF level, for example using the spin-flip philosophy\cite{Mayhall2014} or using single excitations in NOCIS.\cite{Oosterbaan2018,Oosterbaan2019}
%However, methods based on orthogonal excitations struggle to locate symmetry-broken determinants and are often restricted to partially-optimised HF states to prevent determinants from collapsing onto each other.\cite{Mayhall2014}
%Alternatively, symmetry-pure and symmetry-broken determinants HF states can be located using the effectively black-box SCF metadynamics approach.\cite{Thom2008}
%In SCF metadynamics, new initial guesses are obtained from a given stationary state by randomly mixing occupied and virtual MOs, while a biasing potential is used to prevent re-convergence onto previously identified solutions.
%The ability to locate symmetry-broken determinants has made SCF metadynamics the method of choice when applying NOCI to study statically correlated potential energy surfaces.\cite{Thom2009,Jensen2018,Burton2019c}

%%%%%%%%%%%%%%%%%%%%%%%%%%%%%%%%%%%%%%%%%%%%%%%%%%%%%%
\section{Background Theory}
\label{sec:BackgroundTheory}
%%%%%%%%%%%%%%%%%%%%%%%%%%%%%%%%%%%%%%%%%%%%%%%%%%%%%%
%%%%%%%%%%%%%%%%%%%%%%%%%%%%%%%%%%%%%%%%%%%%%%%%%%%%%%
\subsection{Nonorthogonal Configuration Interaction}
\label{subsec:NOCI}
%%%%%%%%%%%%%%%%%%%%%%%%%%%%%%%%%%%%%%%%%%%%%%%%%%%%%%
A multireference NOCI wave function for a given state $k$ is constructed from a linear combination of $\Ndet$ multiple HF determinants as 
\begin{equation}
\ket{\Wfn^{(0)}_{k}} = \ket{\wNOCI_k} = \sum_{x}^{\Ndet} \ket{\wSCF^{x}} \xNOCI_{x k},
\label{eq:NOCIWavefunction}
\end{equation}
where each basis state $\ket{\wSCF^{x}}$ corresponds to a single Slater determinant constructed from a bespoke set of $\Ne$ occupied MOs $\{ \ket{\wMO^{x}_{i}} \}$.
The molecular orbitals (MOs) for each determinant are expanded as a linear combination of $2\Nbas$ (real nonorthogonal) atomic spin orbitals $\{ \ket{\wAO_{\mu}} \}$ as
\begin{equation}
	\ket{\wMO^{x}_{i}} = \sum_{\mu}^{2\Nbas} \ket{\wAO_{\mu}} \up{x}\xMO^{\mu \cdot}_{\cdot i},
\end{equation}
where we have employed the nonorthogonal tensor notation of Head-Gordon \etal{} (see Ref.~\onlinecite{HeadGordon1998} for details).
Optimal NOCI wave functions can then be identified by solving the generalised eigenvalue problem 
\begin{equation}
	\sum_{x}^{\Ndet} \qty( H^{wx} - E S^{wx} ) \xNOCI_{x k} = 0, 
	\label{eq:NOCIeigenvalue}
\end{equation}
where $H^{wx} = \braket{\wSCF^{w}}{\hH}{\wSCF^{x}}$ and $S^{wx} = \brkt{\wSCF^{w}}{\wSCF^{x}}$ are the Hamiltonian and overlap matrices between the nonorthogonal determinants.
These matrix elements can be computed by first exploiting L\"{o}wdin's pairing approach\cite{Amos1961,Lowdin1962} to construct a biorthogonal set of orbitals, and then applying the generalised Slater--Condon rules.\cite{MayerBook}
For a given set of reference HF determinants, the corresponding NOCI wave functions $\{ \ket{\wNOCI_{k}} \}$ can provide state-specific approximations to both ground- and excited-states.\cite{Sundstrom2014,Burton2019c}

%%%%%%%%%%%%%%%%%%%%%%%%%%%%%%%%%%%%%%%%%%%%%%%%%%%%%%
\subsection{Holomorphic Hartree--Fock}
\label{subsec:HoloHF}
%%%%%%%%%%%%%%%%%%%%%%%%%%%%%%%%%%%%%%%%%%%%%%%%%%%%%%

Until recently, the application of NOCI to molecular energy surfaces has been restricted by the disappearance of HF solutions as the molecular structure changes, leading to discontinuities in the NOCI energy.\cite{Thom2009}
To remove these discontinuities, we have developed a modification to HF theory --- known as holomorphic Hartree--Fock (h-HF) --- where real HF states are analytically extended across all molecular geometries.\cite{Hiscock2014, Burton2016, Burton2018, Burton2019c, BurtonThesis}
In the h-HF approach, the real HF equations are analytically-continued into the complex plane by removing the complex-conjugation of orbital coefficients from the HF energy functional and defining the complex-valued h-HF energy as\cite{Hiscock2014,Burton2016}
\begin{equation}
	E_\text{h-HF} = \frac{ \braket{\wSCF^{*}}{\hH}{\wSCF}}{\brkt{\wSCF^{*}}{\wSCF}}
\end{equation}
All real HF solutions remain stationary points of the h-HF energy but, when real HF states coalesce and disappear, their h-HF counterparts continue to exist with complex-valued orbital coefficients.\cite{Burton2016,Burton2018}
As a result, multiple h-HF solutions provide a consistent basis for NOCI across all geometries, allowing smooth and continuous NOCI energy surfaces to be computed.\cite{Burton2019c}

Notably, complex-valued h-HF states are not stationary states of the real HF energy and can have complex-valued non-variational holomorphic energies.
Furthermore, the h-HF wave function is complex-orthogonal,\cite{Burton2016} although unitary normalisation can be recovered by orthonormalising the occupied orbitals.
Beyond NOCI, h-HF theory has provided wider insights into the nature of multiple HF solutions.
For example, it has allowed the identification of new non-Hermitian symmetry conditions for real energies in HF methods\cite{Burton2019b} and has revealed that discrete HF states can be connected as one continuous structure \textit{via} a complex adiabatic connection.\cite{Burton2019a}

%%%%%%%%%%%%%%%%%%%%%%%%%%%%%%%%%%%%%%%%%%%%%%%%%%%%%%
\subsection{Perturbation Theory}
\label{subsec:GenPerturbTheor}
%%%%%%%%%%%%%%%%%%%%%%%%%%%%%%%%%%%%%%%%%%%%%%%%%%%%%%
Perturbation theories partition the exact Hamiltonian as
\begin{equation}
\hH = \Href + \lambda \Hpert,
\end{equation}
where $\Href$ defines the reference Hamiltonian and $\Hpert$ provides the perturbation operator.
The exact wave function $\ket{\Wfn}$ and its energy $E$ are expanded in powers of $\lambda$ as
\begin{equation}
	\ket{\Wfn} = \sum_{n=0}^{\infty} \lambda^{n} \ket{\Wfn^{(n)}} 
	\quad \text{and} \quad
	E              = \sum_{n=0}^{\infty} \lambda^{n} E^{(n)} .
\end{equation}
Following the derivation outlined by Fulde in Ref.~\onlinecite{FuldeBook}, the perturbative corrections $\ket{\Wfn^{(n)}}$ and $E^{(n)}$ can be identified by partitioning the full Hilbert space into a ``model'' space corresponding to the zeroth-order wave function and the remaining ``external'' space.
Projectors onto the model and external spaces can be defined as $\cP$ and $\cQ$ respectively, where 
\begin{equation}
\cP = \ket{\Wfn^{(0)}}\bra{\Wfn^{(0)}} \quad\text{and}\quad \cI = \cP + \cQ.
\end{equation}
Assuming intermediate normalisation $\brkt{\Wfn^{(0)}}{\Wfn}=1$, the operator $\cP$ projects the exact wave function onto the reference wave function, \ie\ $\ket{\Wfn^{(0)}} = \cP \ket{\Wfn}$. 

Decomposing $\hH$ into the model and perturbation spaces leads to the L\"{o}wdin partitioning\cite{Lowdin1962a}
\begin{equation}
\begin{pmatrix}
\cP \hH \cP 	&	\cP \hH \cQ
\\ 
\cQ \hH \cP 	& \cQ \hH \cQ
\end{pmatrix}
\begin{pmatrix}
\cP \ket{\Wfn}
\\ 
\cQ \ket{\Wfn}
\end{pmatrix}
= 
E
\begin{pmatrix}
\cP \ket{\Wfn}
\\ 
\cQ \ket{\Wfn}
\end{pmatrix}.
\label{eq:PartitionedEigenvalue}
\end{equation}
In theory, $\cQ \ket{\Wfn}$ can be eliminated through the substitution
\begin{equation}
	\cQ \ket{\Wfn} = - \big( \cQ (\hH - E) \cQ  \big)^{-1} \cQ \Hpert \cP \ket{\Wfn}
\end{equation}
to construct an effective Hamiltonian $\Heff$ as
\begin{equation}
\Heff = \cP \hH \cP -  \cP \Hpert \cQ \big( \cQ (\hH - E) \cQ  \big)^{-1}  \cQ \Hpert \cP.
\label{eq:Heff}
\end{equation}
where the relationship $\cQ \Href \cP = \cP \Href \cQ = 0$ has been introduced.
However, the presence of the unknown $E$ in Eq.~\eqref{eq:Heff} prevents the eigenvalues of $\Heff$ from being solved exactly. 
Instead, Rayleigh--Schr\"{o}dinger perturbation theory provides an approximate solution by introducing the wave function corrections\cite{FuldeBook}
\begin{equation}
\ket{\Wfn^{(n)}} = - \cX^{-1} (\cY \cX^{-1})^{(n-1)} \cQ \Hpert \cP \ket{\Wfn^{(0)}},
\end{equation}
where  $\cX = \cQ( \Href - E^{(0)}) \cQ$ and $\cY= \cQ (E - E^{(0)} - \Hpert) \cQ$.
Similarly, the energy corrections are given by
\begin{subequations}
\begin{align}
	E^{(0)} &= \braket{\Wfn^{(0)}}{\Href}{\Wfn^{(0)}}
	\\
	E^{(1)} &= \braket{\Wfn^{(0)}}{\Hpert}{\Wfn^{(0)}}
	\\
	E^{(n)} &= \braket{\Wfn^{(0)}}{\Hpert \cQ}{\Wfn^{(n-1)}} \quad \text{for}  \quad n \geq 2.
	\label{eq:E_nth_correction}
\end{align}
\end{subequations}

Eq.~\eqref{eq:E_nth_correction} demonstrates that the second-order energy correction only requires the first-order wave function.
The reference Hamiltonian can sometimes be directly inverted to give
\begin{equation}
	\ket{\Wfn^{(1)}} = - \big( \cQ (\Href - E^{{0}}) \cQ  \big)^{-1} \cQ \Hpert \cP \ket{\Wfn^{(0)}}.
\end{equation}
Alternatively, a variational estimate for $E^{(2)}$ can be identified by minimising the Hylleraas functional\cite{MayerBook} 
\begin{equation}
\begin{split}
\cL[\Wfn^{(1)}] &= \braket{\Wfn^{(1)}}{\pExt (\Href - E^{(0)}) \pExt }{\Wfn^{(1)}} 
\\
&+ \braket{\Wfn^{(1)}}{\pExt \Hpert \pModel}{\Wfn^{(0)}} + \braket{\Wfn^{(0)}}{\pModel \Hpert \pExt}{\Wfn^{(1)}},
\end{split}
\label{eq:ch8:HylleraasFunctional}
\end{equation}
where now $\ket{\Wfn^{(1)}}$ provides an approximation to the exact first-order wave function.
Significantly, this variational procedure allows the second-order energy to be estimated using \emph{any} first-order wave function \ansatz.

%%%%%%%%%%%%%%%%%%%%%%%%%%%%%%%%%%%%%%%%%%%%%%%%%%%%%%
\section{Theoretical Development}
\label{sec:TheoryDevelopment}
%%%%%%%%%%%%%%%%%%%%%%%%%%%%%%%%%%%%%%%%%%%%%%%%%%%%%%
%%%%%%%%%%%%%%%%%%%%%%%%%%%%%%%%%%%%%%%%%%%%%%%%%%%%%%
%\subsection{Challenges of NOCI Perturbation Theory}
%%%%%%%%%%%%%%%%%%%%%%%%%%%%%%%%%%%%%%%%%%%%%%%%%%%%%%
% REMINDER ON NOCI AND SUMMARY
To construct a rigorous second-order diagonalise-then-perturb NOCI correction, we will consider the most general form of a NOCI wave function where each basis state $\ket{\wSCF^{x}}$ may be a complex-valued h-HF determinant and may not correspond to a stationary solution of the real HF equations.
We therefore assume that these basis states do not satisfy Brillouin's condition\cite{SzaboBook} and may not be normalised.
Normalisation can always be recovered without changing the NOCI energy by orthonormalising the occupied orbitals and constructing an orthogonal set of virtual orbitals.
Formally defining a NOCI second-order perturbative correction then requires three key components:
\begin{enumerate}[label=(\roman*),itemsep=0em]
	\item Construction of a simple and well-defined reference Hamiltonian $\Href$;
	\item Identification of the first-order wave function;
	\item Avoidance of intruder state singularities.
\end{enumerate} 
Ideally, any NOCI-based perturbation theory should reduce to either MP2 for a single determinant NOCI wave function, or a multireference approach such as CASPT2 for an equivalent set of orthogonal reference determinants.
However, the nonorthogonal structure of the NOCI reference wave function makes deriving a second-order perturbative correction more challenging than orthogonal approaches such as MP2 or CASPT2.
In this section, we describe how these challenges can be overcome to derive a rigorous second-order perturbative correction that we call NOCI-PT2 theory.
%Furthermore, it is highly desirable that any NOCI perturbation theory should reduce to the conventional single reference MP2 approach in the limit of one determinant, or CASPT2 if a suitable set of mutually orthogonal NOCI determinants is chosen.
%While the NOCI-MP2 approach\cite{Yost2016} satisfies the first of these conditions and reduces to single determinant MP2 theory, it does not reduce to any current multireference theory when orthogonal determinants are used.

%%%%%%%%%%%%%%%%%%%%%%%%%%%%%%%%%%%%%%%%%%%%%%%%%%%%%
\subsection{Defining the Reference Hamiltonian}
\label{subsec:ch8:ReferenceHamiltonian}
%%%%%%%%%%%%%%%%%%%%%%%%%%%%%%%%%%%%%%%%%%%%%%%%%%%%%
% DEFINITION OF REFERENCE HAMILTONIAN
The reference Hamiltonian $\Href$ for single-reference wave functions is generally defined using the one-electron Fock operator in the M\o{}ller--Plesset (MP) partitioning.\cite{Moller1934}
However, for multiconfigurational reference wave functions, there is no universal set of occupied MOs and the Fock operator definition becomes ambiguous.
Instead, a generalised Fock operator $\hFg$ can be defined in the atomic spin-orbital basis as\cite{Andersson1990,Andersson1992}
\begin{equation}
\Fg \qty[\bPg]_{\mu \nu} = \oneE_{\mu \nu} + \sum_{\sigma \tau}^{2\Nbas} \twoEas{\mu}{\sigma}{\nu}{\tau} (\Pg)^{\tau \sigma},
\label{eq:ch8:genFock}
\end{equation}
where $\bPg$ is the one-particle reduced density matrix of the multireference wave function, $h_{\mu \nu}$ are the one-electron integrals, and $\twoEas{\mu}{\sigma}{\nu}{\tau}$ are the antisymmetrised two-electron integrals.\cite{HelgakerBook}
For a single determinant reference wave function, this generalised Fock operator reduces to the MP zeroth-order Hamiltonian since the MOs become eigenfunctions of $\Fg$.
However, to ensure that a multiconfigurational wave function remains an eigenfunction of $\Href$, the generalised Fock operator must be further modified using the model and external space projectors to give\cite{YarkonyBook}
\begin{equation}
	\Href = \pModel \hFg\pModel + \pExt \hFg \pExt.
	\label{eq:ch8:zerothH}
\end{equation}
%For example, including these projectors for a CASSCF wave function leads to the CASPT2 zeroth-order Hamiltonian defined in Eq.~\eqref{eq:ch1:HrefModel}.

% NOCI REFERENCE HAMILTONIAN
Like all multireference wave functions, the target NOCI wave function $\ket{\wNOCI_{k}}$ also lacks a well-defined set of occupied orbitals.
Furthermore, since each NOCI basis state is an independently optimised Slater determinant, there is no common set of MOs to build a model reference Hamiltonian.
Instead, a state-specific generalised Fock operator built from the NOCI wave function $\ket{\wNOCI_{k}}$ can be combined with the projectors 
\begin{subequations}
\begin{align}
	\pModel &= \ket{\wNOCI_{k}}\bra{\wNOCI_{k}}
    \label{eq:NOCI_P}
    \\
    \pExt  &= \cI - \ket{\wNOCI_{k}}\bra{\wNOCI_{k}}
    \label{eq:NOCI_Q}
\end{align}
\end{subequations}
to define a zeroth-order Hamiltonian of the form given in Eq.~\eqref{eq:ch8:zerothH}.
These projectors, and the generalised Fock operator, depend on the target (ground- or excited-state) NOCI wave function $\ket{\wNOCI_{k}}$.
The resulting perturbation theory is therefore explicitly state specific, and we will henceforth drop the index $k$.
Crucially, this reference Hamiltonian reduces to the MP zeroth-order Hamiltonian for a single determinant wave function, and is closely related to the CASPT2 zeroth-order Hamiltonian when the NOCI basis states span an orthogonal complete active space.
Furthermore, since we always include spin-flipped pairs of UHF solutions in our NOCI wave function, the generalised Fock operator is unambiguously defined in a  spin-averaged form by construction.

%%%%%%%%%%%%%%%%%%%%%%%%%%%%%%%%%%%%%%%%%%%%%%%%%%%%%
\subsection{Identifying the First-Order Interacting Space}
\label{subsec:ch8:ExternalSpace}
%%%%%%%%%%%%%%%%%%%%%%%%%%%%%%%%%%%%%%%%%%%%%%%%%%%%%

% DEFINITION OF THE EXTERNAL SPACE
For orthogonal perturbation theories, an \ansatz{} for the first-order wave function can be built from the set of external-space determinants that couple to the reference wave function through $\hH$, defining the `first-order interacting space'.\cite{YarkonyBook}
In contrast, partitioning the Hilbert space determinants into model and external spaces is impossible for NOCI since every Hilbert space determinant can have a component in the reference wave function.
This absence of a well-defined Hilbert space partitioning is shared with other reference wave functions such as matrix product states used in the density matrix renormalisation group approach.\cite{Sharma2018,Guo2018,Guo2018a}
As a result, while the model space projector $\pModel$ can still be defined using Eq.~\eqref{eq:NOCI_P}, the external space projector $\pExt$ cannot be constructed explicitly.
%In particular, since each reference NOCI determinant is build from a different set of orbitals, there is no well-defined resolution of the identity $\cI$ to expand $\cQ$ using Eq.~\eqref{eq:NOCI_Q}.
Instead, we proceed with only the representation $\pExt = \cI - \pModel$ and identify $\ket{\Wfn^{(1)}}$ by solving the first-order wave function equation
\begin{equation}
\pExt (\Href - E^{(0)}) \pExt  \ket{\Wfn^{(1)}} = - \pExt \Hpert \ket{\Wfn^{(0)}},
\label{eq:ch8:firstOrderEq}
\end{equation}
where $E^{(0)} = \braket{\wNOCI}{\Href}{\wNOCI}$.

In orthogonal perturbation theory, $\ket{\Wfn^{(1)}}$ can be expanded using the single and double excitations from each reference determinant in the multideterminantal reference wave function.\cite{Andersson1990,Andersson1992}
However, since the NOCI wave function can span the whole Hilbert space, expanding the first-order wave function in terms of excited determinants built from a \emph{single} set of orthogonal MOs requires excitations of all orders.
Instead, a first-order wave function \ansatz{} can be built by considering the combined Hilbert spaces constructed from every reference determinant in the NOCI wave function.
While this approach requires multiple representations of the full Hilbert space, only single and double excitations from each reference determinant provide unique contributions to the second-order energy.

Considering the $\Href$ from Eq.~\eqref{eq:ch8:zerothH} and the expansion 
\begin{equation}
\begin{split}
\pExt \Hpert \ket{\Wfn^{(0)}} 
&= (\cI - \pModel) (\hH - \Href) \ket{\Wfn^{(0)}}  
\\
&= (\hH - \Href - \pModel \hH + \pModel \Href)  \ket{\Wfn^{(0)}}
\\
&= (\hH - E_{\text{ref}})  \ket{\Wfn^{(0)}},
\label{eq:QVsplit}
\end{split}
\end{equation}
the second-order energy is given by
\begin{equation}
E^{(2)} = \braket{\Wfn^{(1)}}{\pExt \Hpert}{\Wfn^{(0)}} 
= \braket{\Wfn^{(1)}}{\hH - E_{\text{ref}}}{\Wfn^{(0)}}, 
\label{eq:SecondOrderEnergy}
\end{equation}
where $E_{\text{ref}} = \braket{\Wfn^{(0)}}{\hH}{\Wfn^{(0)}} = E^{(0)} + E^{(1)}$.
%The first-order wave function \ansatz{} may not be orthogonal to $\ket{\Wfn^{(0)}}$, although the presence of $\cQ$ in Eq.~\eqref{eq:SecondOrderEnergy} removes any components of $\ket{\Wfn^{(1)}}$ that are already present in the model space.
Inserting the NOCI wave function \eqref{eq:NOCIWavefunction} and expanding the action of $\hH$ on each determinant then gives
\begin{equation}
\begin{split}
E^{(2)}
&= \sum_{w}^{\Nref} \braket{\Wfn^{(1)}}{\hH- E_{\text{ref}}}{\wSCF^{w0}}  \xNOCI_{w}
\\
&= \sum_{w}^{\Nref} 
\Bigg[ 
\sum_{I \in \cW}^{\mathcal{SD}} \brkt{\Wfn^{(1)}}{\wSCF^{wI}} H^{wI,w0} 
\\
&\quad\quad\quad\quad\, + \brkt{\Wfn^{(1)}}{\wSCF^{w0}} \qty( H^{w0,w0} -E_{\text{ref}}) \Bigg] \xNOCI_{w},
\end{split}
\label{eq:FirstOrderWorkingEq3}
\end{equation} 
where $\ket{\wSCF^{wI}}$ denotes an excitation from the reference determinant $\ket{\wSCF^{w0}} \equiv \ket{\wSCF^{w}}$ and $\sum_{I \in \cW}^{\mathcal{SD}}$ indicates the sum over single and double excitations in the Hilbert space $\cW$ built from determinant $w$.
Crucially, triple (or higher) replacement determinants in one Hilbert space representation can only contribute to $E^{(2)}$ if they correspond to a linear combination of single and double excitations from other reference wave functions.
The first-order interacting space, containing determinants that provide a unique contribution to the second-order energy, therefore includes only single and double excitations from \emph{each} reference determinant in the NOCI expansion.
Strict orthogonality with the reference wave function can be ensured by constructing the expansion basis from the projected determinants
\begin{equation}
	\ket{\Omega^{wI}} = \pExt  \ket{\wSCF^{wI}}.
	\label{eq:FOIS}
\end{equation}
The first-order wave function can then be expanded as
\begin{equation}
\ket{\Wfn^{(1)}} = \sum_{w}^{\Nref} \sum_{I \in \cW}^{\mathcal{SD}} \ket{\Omega^{wI}} \xFO_{wI},
\label{eq:firstOrderWfn}
\end{equation}
where $\xFO_{w I}$ represents variable coefficients that remain to be identified.
Note that the reference determinants are not included in the first-order interacting space as they satisfy the NOCI eigenvalue equation \eqref{eq:NOCIeigenvalue} and do not contribute to the second-order energy.

%%%%%%%%%%%%%%%%%%%%%%%%%%%%%%%%%%%%%%%%%%%%%%%%%%%%%%
\subsection{Computing the Second-Order Energy}
\label{subsec:ch8:NOCIPT2Equations}
%%%%%%%%%%%%%%%%%%%%%%%%%%%%%%%%%%%%%%%%%%%%%%%%%%%%%%
Using Eq.~\eqref{eq:firstOrderWfn} and the idempotency of $\pExt$, the optimal coefficients $\xFO_{w I}$ can be identified by projecting the first-order perturbation equation \eqref{eq:ch8:firstOrderEq} onto the first-order interacting space to give the NOCI-PT2 equations 
%\begin{equation}
%\sum_{w}^{\Nref} \sum_{I \in \cW}^{\mathcal{SD}} 
%\pExt (\Href - E^{(0)}) \pExt \ket{\wSCF^{wI}} \xFO_{w I}
%= - \pExt \Hpert \ket{\Wfn^{(0)}}.
%label{eq:ch8:FirstOrderWorkingEq4}
%\end{equation}
%The optimal coefficients $\xFO_{w I}$ can then be identified by projecting this expression onto the first-order interacting space to give the NOCI-PT2 equations
\begin{equation}
\begin{split}
\sum_{w}^{\Nref} \sum_{I \in \cW}^{\mathcal{SD}} \bra{\Omega^{xJ}} \Href - &E^{(0)} \ket{\Omega^{wI}}  \xFO_{w I}
= - \braket{\Omega^{xJ}}{ \Hpert}{\Wfn^{(0)}}.
\end{split}
\label{eq:ch8:FirstOrderProj}
\end{equation}
In principle, higher excitations can couple to the single and double excitations through $(\Href - E^{(0)})$ and alter the  expansion coefficients of the first-order wave function. 
However, the approximate first-order wave function obtained from Eq.~\eqref{eq:ch8:FirstOrderProj} provides a variational estimate of the $E^{(2)}$ through the Hylleraas functional \eqref{eq:ch8:HylleraasFunctional}.
The exact second-order energy would be obtained by including excitations of every order in the first-order wave function equation.

Introducing the matrix elements
\begin{subequations}
	\begin{align}
	F^{xJ, wI} &=  \braket{\Omega^{xJ}}{\Href}{\Omega^{wI}} =\braket{\wSCF^{xJ}}{\pExt \Href \pExt}{\wSCF^{wI}},
	\label{eq:NOCIPT2_fmat}
	\\
	Q^{xJ, wI} &= \brkt{\Omega^{xJ}}{\Omega^{wI}} = \braket{\wSCF^{xJ}}{\pExt}{\wSCF^{wI}},
	\label{eq:NOCIPT2_smat}
	\\
	V^{xJ} &= \braket{\Omega^{xJ}}{\Hpert}{\Wfn^{(0)}}	= \braket{\wSCF^{xJ}}{\hH- E_\text{ref}}{\Wfn^{(0)}},
	\label{eq:NOCIPT2_vec}
	\end{align}
	\label{eq:NOCIPT2}
\end{subequations}
where Eq.~\eqref{eq:NOCIPT2_vec} has exploited the expansion \eqref{eq:QVsplit},
allows the NOCI-PT2 equations to be  expressed as the linear equation 
\begin{equation}
\big(\bF - E^{(0)} \bQ \big) \ba = - \bV.
\label{eq:ch8:FirstOrderMat}
\end{equation}
The corresponding second-order energy is then given by
\begin{equation}
E^{(2)} = \ba^{\dagger} \, \bV.
\end{equation}
Explicitly expanding $\pExt = \cI - \ket{\Wfn^{(0)}}\bra{\Wfn^{(0)}}$ and noting that $\pExt \Href \pExt = \pExt \hFg \pExt$ allows the matrix elements \eqref{eq:NOCIPT2_fmat} and \eqref{eq:NOCIPT2_smat} to be expressed as
\begin{subequations}
\begin{align}
\hspace{-.7em}F^{xJ, wI} =\, &\braket{\wSCF^{xJ}}{\hFg}{\wSCF^{wI}} + E^{(0)} \brkt{\wSCF^{xJ}}{\Wfn^{(0)}}\brkt{\Wfn^{(0)}}{\wSCF^{wI}} \nonumber
\\
&- \braket{\wSCF^{xJ}}{\hFg}{\Wfn^{(0)}}\brkt{\Wfn^{(0)}}{\wSCF^{wI}} \nonumber
\\
&- \brkt{\wSCF^{xJ}}{\Wfn^{(0)}}\braket{\Wfn^{(0)}}{\hFg}{\wSCF^{wI}},
\label{eq:ch8:FmatExpanded}
\\[1em]
Q^{xJ, wI} =\, &\brkt{\wSCF^{xJ}}{\wSCF^{wI}} - \brkt{\wSCF^{xJ}}{\Wfn^{(0)}}\brkt{\Wfn^{(0)}}{\wSCF^{wI}}.
\label{eq:ch8:SmatExpanded}
\end{align}
\end{subequations}
respectively.
%In these expansions, terms containing $\Wfn^{(0)}$ account for components of the first-order interacting space that are already present in the reference wave function.

In principle, the NOCI-PT2 equations can be directly solved by inverting the linear equation \eqref{eq:NOCIPT2}.
However, redundancies can exist among the perturbing determinants, particularly since single and double excitations built from different NOCI reference determinants are likely to be linearly dependent.
The presence of these redundancies leads to a null space in the overlap matrix \eqref{eq:NOCIPT2_smat} and must be taken into account by first diagonalising the overlap matrix $\bQ$ to identify the non-null eigenvectors.
Constructing the projection matrix into the non-null space $\bX$, where 
\begin{equation}
\bX^{\dagger} \bQ \bX = \bI,
\label{eq:ch8:Xdef}
\end{equation}
allows the NOCI-PT2 equations to be transformed to the form
\begin{equation}
(\tilde{\bF} - E^{(0)} \bI )\tilde{\ba} = -\tilde{\bV},
\end{equation}
where $\tilde{\bF} = \bX^{\dagger} \bF \bX$, $\tilde{\ba} = \bX^{\dagger} \bQ \ba $ and  $\tilde{\bV} = \bX^{\dagger} \bV$.
Identifying the diagonal matrix 
\begin{equation}
\bDelta = \bY^{\dagger} \tilde{\bF} \bY - E^{(0)} \bI,
\label{eq:ch8:Ydef}
\end{equation} 
where $\bY$ is a transformation matrix built from the eigenvectors of $\tilde{\bF}$,
then allows the the first-order wave function coefficients to be evaluated as
\begin{equation}
\ttilde{\ba} = - \invbDelta \ttilde{\bV},
\end{equation}
where $\ttilde{\ba} = \bY^{\dagger} \tilde{\ba}$ and $\ttilde{ \bV } = \bY^{\dagger} \tilde{\bV}$.
The second-order energy correction can then be computed as 
\begin{equation}
E^{(2)} = \ttilde{\ba}^{\dagger} \ttilde{\bV}  =  - \ttilde{\bV}^{\dagger} \invbDelta\, \ttilde{\bV}.
\end{equation}

In practice, since all single and double excitations from each reference determinant are included in the first-order wave function, the dimensions of the NOCI-PT2 matrices will scale as $\Or(\Nref \Ne^2 \Nbas^2 )$ and exact diagonalisation will generally be intractable.
%Furthermore, in contrast to CASPT2, the NOCI-PT2 matrices do not possess any block-diagonal form that would reduce the cost of exact diagonalisation.
The NOCI-PT2 equations must therefore be evaluated using an iterative linear approach by casting Eq.~\eqref{eq:ch8:FirstOrderMat} into the form
\begin{equation}
\bM \ba = - \bV
\label{eq:ch8:linearProblem}
\end{equation}
where $\bM = \bF - E^{(0)} \bQ$.
Using Eqs.~\eqref{eq:ch8:FmatExpanded} and \eqref{eq:ch8:SmatExpanded} allows the matrix elements of $\bM$ to be explicitly expanded as
\begin{align}
M^{xJ,wI} = &\braket{\wSCF^{xJ}}{\hFg}{\wSCF^{wI}} \nonumber
\\
&- \braket{\wSCF^{xJ}}{\hFg}{\Wfn^{(0)}}\brkt{\Wfn^{(0)}}{\wSCF^{wI}} \label{eq:ch8:Mmatrix}
\\
&- \brkt{\wSCF^{xJ}}{\Wfn^{(0)}}\braket{\Wfn^{(0)}}{\hFg}{\wSCF^{wI}}  \nonumber
\\
&- E^{(0)} \qty[\brkt{\wSCF^{xJ}}{\wSCF^{wI}}  - 2 \brkt{\wSCF^{xJ}}{\Wfn^{(0)}}\brkt{\Wfn^{(0)}}{\wSCF^{wI}}].  \nonumber
\end{align}
Since the redundancy in the perturbing determinants appears in both $\bM$ and $\bV$ in exactly the same way, the NOCI-PT2 linear problem must be consistent and possess a unique solution.
However, the presence of small non-zero eigenvalues in $\bM$ may slow the iterative convergence, although this effect can be alleviated using preconditioning schemes.\cite{VanDerVorstBook}
We find that the restarted generalised minimum residual (GMRES) method with a diagonal preconditioner provides sufficiently well-behaved convergence for the cases studied in this work [see Ref.~\onlinecite{VanDerVorstBook} for more details].

%%%%%%%%%%%%%%%%%%%%%%%%%%%%%%%%%%%%%%%%%%%%%%%%%%%%%%
\subsection{Relationship to CASPT2 and SUPT2}
\label{subsec:CASPT2_SUPT2_Relationship}
%%%%%%%%%%%%%%%%%%%%%%%%%%%%%%%%%%%%%%%%%%%%%%%%%%%%%%
The matrix form of the NOCI-PT2 equations described in Section~\ref{subsec:ch8:NOCIPT2Equations} highlights a close relationship with CASPT2. 
If the NOCI wave function is constructed from a suitable set of orthogonal determinants spanning an active orbital space, then NOCI-PT2 essentially reduces to an uncontracted form of CASPT2.
Note that NOCI-PT2 and CASPT2 are never exactly equivalent since the CASPT2 zeroth-order Hamiltonian involves additional projectors onto the various subspaces defined by the active and inactive sets of orbitals.\cite{YarkonyBook}
Regardless, our NOCI-PT2 method can be regarded as the rigorous nonorthogonal extension of CASPT2, and we should expect to achieve a similar degree of accuracy.

NOCI-PT2 is also closely related to the recently derived Spin-Symmetry-Projected HF Perturbation Theory (SUPT2) that provides a perturbative correction to spin-symmetry-projected HF (SUHF) wave functions.\cite{Tsuchimochi2019}
When the integration of the spin-symmetry projection operator is discretised, the SUHF wave function reduces to a NOCI expansion containing only degenerate Slater determinants.\cite{Jimenez-Hoyos2012}
In SUPT2, the first-order wave function is expanded using contracted symmetry-projections of the single and double excitation determinants (``projection-after-excitation'').\cite{Tsuchimochi2019}
However, our NOCI wave functions are not restricted to only symmetry-projected reference determinants, but can also include non-degenerate determinants representing excited or diabatic states.
A contracted projection-after-excitation first-order wave function is therefore not possible in NOCI-PT2, and thus we use the single and double excitations from each reference determinant (equivalent to ``excitation-after-projection'').
NOCI-PT2 therefore generalises SUPT2 to all multiconfigurational wave functions built from nonorthogonal Slater determinants.

%%%%%%%%%%%%%%%%%%%%%%%%%%%%%%%%%%%%%%%%%%%%%%%%%%%%%%
\subsection{Intruder States and Imaginary Shifts}
\label{subsec:NOCIPT2ImShift}
%%%%%%%%%%%%%%%%%%%%%%%%%%%%%%%%%%%%%%%%%%%%%%%%%%%%%%

% INTRODUCTION TO IMAGINARY SHIFT
Like all perturbation theories, particularly multireference approaches, NOCI-PT2 is susceptible to the effects of intruder states arising from singularities in the second-order perturbation equations. 
Intruder states formally arise when the eigenvalues of $\tilde{\bF}$ (diagonal elements of $\bDelta$) fall to zero, creating poles in $\invbDelta$ that cause the first-order wave function coefficients to blow-up.
While the presence of intruder states often indicates a bad choice of Hamiltonian partitioning or reference wave function, their effects can be mitigated using either a real\cite{Roos1995} or imaginary shift\cite{Forsberg1997} in $\Href$.
Real energy shifts can only move the poles in $\invbDelta$ along the real axis, often  causing the divergences to simply occur at a different molecular geometry.
In contrast, imaginary shifts move the poles into the complex plane and provide a more robust way of removing intruder state divergences.
Furthermore, while both real and imaginary shifts introduce a small distortion to energy surface away from any singularities, this effect is smaller using imaginary shifts.\cite{Forsberg1997}
For these reasons, we will only consider the introduction of an imaginary shift to remove singularities in NOCI-PT2.

% DISCUSSION OF DIFFICULTY
While an imaginary shift formally corresponds to the transformation 
\begin{equation}
	\Href \rightarrow \Href + \I \epsilon \cQ,
	\label{eq:imagShiftH}
\end{equation}
the original imaginary-shifted CASPT2 implementation sought to retain real arithmetic by introducing the modified zeroth-order Hamiltonian $\Href - E^{(0)} = \Href - E^{(0)}  + \epsilon^2 (\Href^{\text{D}} - E^{(0)})^{-1}$, where $\Href^{\text{D}} $ is the diagonal part of $\Href$.\cite{Forsberg1997}
Alternatively, the SUPT2 approach introduced a real-valued first-order wave function equation by back-transforming the imaginary-shifted amplitudes, although this modified expression becomes quadratic in $\bM$.\cite{Tsuchimochi2019}
In the case of NOCI-PT2, both of these approaches require either the inverse overlap matrix or a diagonal approximation for the matrix elements between first-order interacting determinants, as outlined in Appendix~\ref{apdx:ImagShift}.
Our preliminary investigations indicate that these approximations lead to extremely poor iterative convergence.
Instead, since including h-HF states in the NOCI basis already requires complex arithmetic, we find that the explicit imaginary-shifted Hamiltonian \eqref{eq:imagShiftH} provides the fastest and most robust convergence.

% DERIVATION OF EQUATIONS
In our imaginary-shifted NOCI-PT2 approach, we therefore directly introduce the transformation \eqref{eq:imagShiftH} using the modified matrix 
\begin{equation}
    \bsM 
    %= \bF - (E^{(0)} - \I \epsilon) \bQ 
    = \bM + \I\epsilon\, \bQ
 \label{eq:imShift}
\end{equation}
to give the shifted linear problem
\begin{equation}
	\bsM\, \ba = - \bV.
\label{eq:imShiftLinProb} 
\end{equation}
The shifted matrix $\bsM$ is non-Hermitian in general, although the complex GMRES iterative approach is sufficiently flexible to allow such linear problems to solved.\cite{VanDerVorstBook}
Once an approximate first-order wave function has been identified, an upper bound on the second-order energy can be computed by taking the real component of the Hylleraas functional as
\begin{equation}
	E^{(2)} = \Re[ \ba^{\dagger} \bsM \ba + \ba^{\dagger} \bV + \bV^{\dagger} \ba ].
\end{equation}
Using the real component of the Hylleraas functional allows this estimate for $E^{(2)}$ to be evaluated without needing to explicitly evaluate the unshifted matrix $\bM$.

%%%%%%%%%%%%%%%%%%%%%%%%%%%%%%%%%%%%%%%%%%%%%%%%%%%%%%
\section{Computational Details}
\label{sec:ComputationalDetails}
%%%%%%%%%%%%%%%%%%%%%%%%%%%%%%%%%%%%%%%%%%%%%%%%%%%%%%

We have implemented our NOCI-PT2 approach in a developmental version of \textsc{Q-Chem 5.2}\cite{QChem} using the \textsc{LIBNOCI} library reported in Ref.~\onlinecite{Burton2019c}.
At present, we build all matrix elements using the generalised Slater--Condon rules\cite{MayerBook} and explicitly store the NOCI-PT2 matrices.
Due to the scaling of the first-order interacting space, storing these matrices creates a memory bottleneck in our implementation.
In the longer term, accelerating the computation of the NOCI-PT2 matrix elements using the nonorthogonal Wick's theorem\cite{RingBook,Nite2019b} should allow an on-the-fly implementation that will remove this memory bottleneck, and we will investigate this in a future publication.

To iteratively solve the NOCI-PT2 equations, we use the restarted GMRES algorithm with a diagonal preconditioner\cite{VanDerVorstBook} and reset the iterative subspace every 200 iterations.
Convergence of the GMRES algorithm is judged using the root-mean-squared length of the residual vector $\br = \bM \ba + \bV$ (or $\br = \bsM \ba + \bV$) with a threshold value of $10^{-7}$.
In some cases, the presence of redundancies in the first-order interacting space can lead to slow iterative convergence of the NOCI-PT2 linear problem.
In the future, we hope to introduce a contracted scheme that allows the effect of these redundancies to be reduced, and research in this direction is ongoing.

All h-HF and NOCI energies were calculated using the \libnoci{} library in \qchem{}\cite{QChem} following the general approach described in Ref.~\onlinecite{Burton2019c}.
Holomorphic DIIS extrapolation was used to accelerate the h-HF self-consistent-field approach.\cite{Burton2016} 
Where h-HF energies become complex, only the real component is shown.

Benchmark CASSCF and CASPT2 energies were computed in the \openmolcas{} package, \cite{OpenMolcas} while NOCI-MP2 energies were computed using an in-house implementation in a developmental version of the \libnoci{} library in \qchem{}.
All NOCI-MP2 calculations use the size-consistent ``version 2'' approach described in Refs.~\onlinecite{Yost2016} and \onlinecite{Yost2019}.
Exact FCI energies for the square \ce{H4} and \ce{LiF} molecules were computed using the \orca{}\cite{ORCA} and \mrcc{}\footnote{{\scshape MRCC}, a quantum chemical program suite written by M. K\'{a}llay, P. R. Nagy, Z. Rolik, D. Mester, G. Samu, J. Csontos, J. Cs\'{o}ka, B. P. Szab\'{o}, L. Gyevi-Nagy, I. Ladj\'{a}nszki, L. Szegedy, B. Lad\'{o}czki, K. Petrov, M. Farkas, P. D. Mezei, and B. H\'{e}gely. See also Z. Rolik, L. Szegedy, I. Ladj\'{a}nszki, B. Lad\'{o}czki, and M. K\'{a}llay, {\it J. Chem. Phys.} {\bf 139}, 094105 (2013), as well as: \href{www.mrcc.hu}{www.mrcc.hu}} packages respectively.
Near-exact energies for \ce{F2} were computed using the CIPSI selected-CI method implemented in \qp{}\cite{QuantumPackage2.0} and were converged to an extrapolated error less than 0.01~$\mEh$.

%%%%%%%%%%%%%%%%%%%%%%%%%%%%%%%%%%%%%%%%%%%%%%%%%%%%%%
\section{Results and Discussion}
\label{sec:Results}
%%%%%%%%%%%%%%%%%%%%%%%%%%%%%%%%%%%%%%%%%%%%%%%%%%%%%%

The application of NOCI-PT2 to strongly correlated wave functions is first illustrated using the symmetric stretch of the square \ce{H4} molecule, which possesses a degenerate RHF ground-state across all geometries.
We then consider the ground-state dissociation of \ce{F2} as a molecular example exhibiting strong static and dynamic correlation.
Finally, we  assess the performance of NOCI-PT2 for predicting the ionic-neutral avoided crossing in the \ce{LiF} binding curve.
In each case, NOCI-PT2 is compared to CASPT2 using an equivalent active space and, where possible, NOCI-MP2 computed with the same NOCI reference wave function.
Energies are provided in atomic units of Hartrees ($\Eh$) throughout.

\subsection{Square \ce{H4}: Symmetric Stretch}
\label{subsec:ch8:H4}
%%%%%%%%%%%%%%%%%%%%%%%%%%%%%%%%%%%%%%%%%%%%%%%%%%%%%%

%============================================================%
\begin{figure*}[tb!]
	\includegraphics[width=\linewidth]{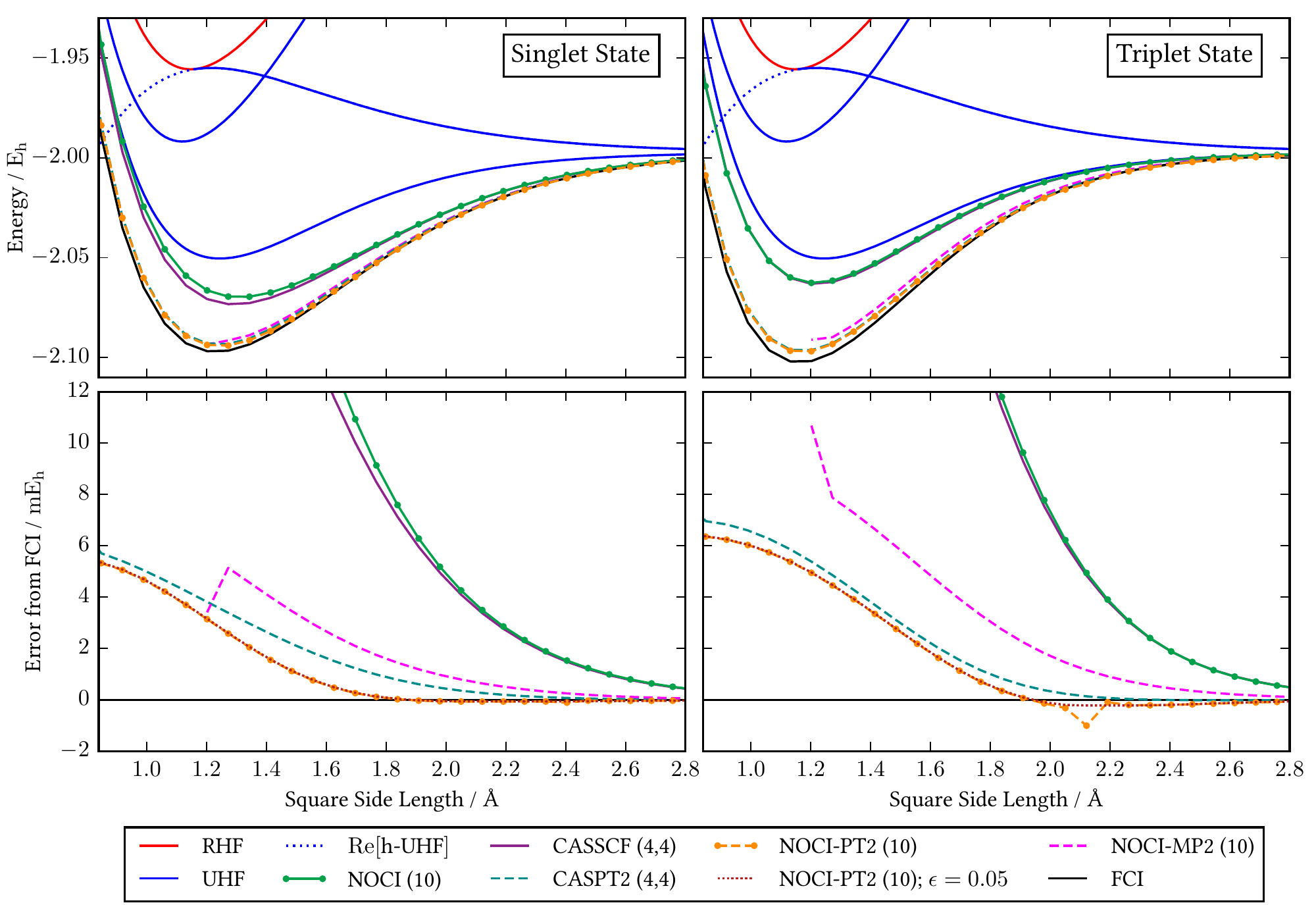}
	\caption[Singlet NOCI-PT2 binding curve for  symmetric stretch of square \ce{H4}]
	{   \label{fig:h4_cc-pvdz}
		\textit{Top:} NOCI~(10) and NOCI-PT2~(10) energies for the lowest-energy singlet (left) and triplet (right) states along the symmetric stretch of square \ce{H4} (cc-pVDZ).
		The NOCI basis comprises the two degenerate ground RHF states and eight (h-)UHF solutions with degeneracies of two, four, and two in order of ascending energy at the dissociation limit.
		\textit{Bottom:} Error in the singlet (left) and triplet (right) NOCI-PT2 energy with respect to the exact FCI energy compared to NOCI-MP2 using the same reference NOCI~(10) wave functions and CASPT2~(4,4).
		An imaginary shift of $\epsilon = 0.05~\Eh$ completely removes the kinks in the NOCI-PT2 energies while leaving the rest of the binding curve unchanged.
	}
\end{figure*}
%============================================================%

% INTRODUCTION AND REFERENCE CALCULATIONS
To demonstrate the performance of NOCI-PT2, we first consider the symmetric stretch of the square \ce{H4} molecule.
Square \ce{H4} represents an interesting test for electronic structure methods as the ground RHF state is degenerate across all geometries, leading to a low-lying singlet and triplet state.
In addition, HF symmetry-breaking relative to the RHF ground state occurs along the full binding curve,\cite{Burton2016} indicating the presence of static correlation in the exact wave function.
Since the symmetric stretch dissociates to four one-electron \ce{H} atoms, the energy of the symmetry-broken UHF states is exact at infinite separation.
As a result, the electron correlation varies from strong static effects in the dissociation limit, to a combination of static and dynamic correlation in the equilibrium regime.

% HF AND NOCI
Using the cc-pVDZ basis,\cite{Dunning1989} we locate ten low-lying HF states that provide a dominant contribution to a NOCI wave function, with degeneracies of two, four, two and two in order of ascending energy in the dissociation limit as shown in Fig.~\ref{fig:h4_cc-pvdz}.
Following these states across the full symmetric stretch, we find that the four-fold degenerate UHF states (solid blue) coalesce with the RHF ground-states (red solid) at two degenerate Coulson--Fischer points (square side length $\approx \SI{1.15}{\angstrom}$).
At shorter bond lengths, two degenerate pairs of h-UHF states continue to exist with complex orbital coefficients and energies related by complex conjugation.
The combined set of h-UHF states and real HF states creates a ten-dimensional basis for NOCI --- henceforth denoted NOCI~(10) ---  allowing the ground-state singlet state (top-left panel) and triplet state (top-right panel) to be computed across all geometries, as shown in Fig.~\ref{fig:h4_cc-pvdz}.
The NOCI~(10) expansions provide the exact energies in dissociation, but are missing a significant amount of the correlation energy for shorter bond lengths.
However, NOCI~(10)  provides a similar binding curve to CASSCF~(4,4) with an active space containing the four lowest-energy MOs, indicating that NOCI~(10)  is successfully capturing static correlation effects across all geometries.
The remaining error in the NOCI~(10) energy near equilibrium can therefore be attributed to dynamic correlation effects.

%============================================================%
\begin{figure*}[tb!]
	\includegraphics[width=\linewidth]{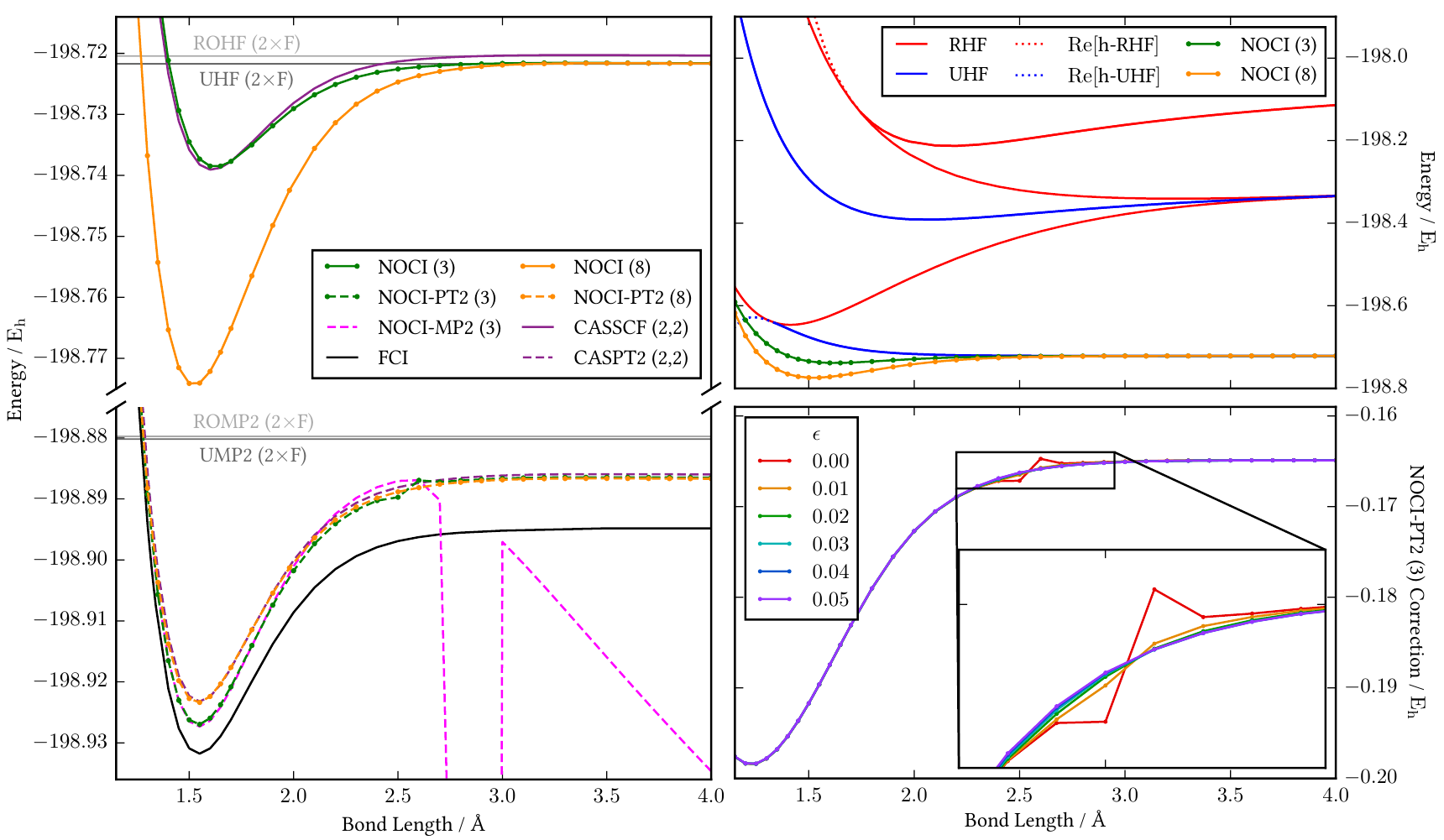}
	\caption[NOCI-PT2 binding curve for the fluorine dimer]
	{   \label{fig:f2_6-31g}
		\textit{Left:} Absolute NOCI and NOCI-PT2 energies for the \ce{F2}  binding curve (6-31G) using the NOCI~(3) and NOCI~(8) reference wave functions.
		\textit{Right Top:} The eight (h-)HF solutions used to define the NOCI expansion basis, with degeneracies of two, one, two, one, and two in order of ascending energy in the dissociation limit.
		\textit{Right Bottom:} Comparison of imaginary shifts for the NOCI-PT2~(3) perturbative correction, demonstrating the systematic removal of the intruder-state singularity while the remainder of the binding curve is left unchanged. 
		The magnitude of the imaginary shift $\epsilon$ is provided in units of Hartrees ($\Eh$).
	}
\end{figure*}
%============================================================%

% PERTURBATIVE CORRECTIONS
Adding the NOCI-PT2 correction on top of the NOCI~(10) wave function provides a significant improvement to both the singlet and triplet energies.
In the equilibrium region, the error in the NOCI-PT2 energy falls below 6~$\mEh$ for both states, providing a marginally better estimate than CASPT2 using the CASSCF~(4,4) reference wave function (bottom panels in Fig.~\ref{fig:h4_cc-pvdz}).
The improvement in the NOCI-PT2 energy relative to CASPT2 is particularly notable since the CASSCF reference wave function has a slightly lower variational energy than NOCI (10) in both cases.
Furthermore, NOCI-PT2 provides a significantly improved estimate of the equilibrium geometry compared to the reference NOCI energy for both the singlet and triplet binding curves.
In contrast, adding the NOCI-MP2 correction to the same NOCI~(10) reference wave functions provides the worst perturbative correction across all geometries, and breaks down completely in the vicinity of the Coulson--Fischer point and at shorter bond lengths.
This breakdown of NOCI-MP2 arises from the emergence of unphysical negative eigenvalues in the NOCI-MP2 overlap matrix that we believe are caused by ignoring the overlap of the perturbed wave functions in the exact matrix elements [see Refs.~\onlinecite{Yost2013}~and~\onlinecite{Yost2016}].

% ADDITION OF AN IMAGINARY SHIFT
While the CASPT2 energy is smooth across all geometries, both the singlet and triplet NOCI-PT2 energies exhibit a small ``kink'' in the binding curve corresponding to a singularity in the NOCI-PT2 equations.
These kinks are visible in the NOCI-PT2 energy at square side lengths of around $\SI{2.4}{\angstrom}$ and $\SI{2.2}{\angstrom}$ for the singlet and triplet states respectively.
Adding a small imaginary shift of $\epsilon = 0.05~\Eh$ as described in Section~\ref{subsec:NOCIPT2ImShift} allows the effect of the singularities to be entirely removed, while leaving the NOCI-PT2 energy virtually unchanged for the remainder of the binding curve.
Notably, the imaginary shift required to mitigate the singularities appears to be an order of magnitude smaller than the recommended values used in other multireference perturbation theories.\cite{Forsberg1997,Chang2012}
We believe that this reduction is the result of using the direct complex implementation of the imaginary shift rather than relying on approximations to retain real arithmetic.
%Our imaginary shift implementation is therefore a promising approach for handling singularities in the NOCI-PT2 energy.

%%%%%%%%%%%%%%%%%%%%%%%%%%%%%%%%%%%%%%%%%%%%%%%%%%%%%%
\subsection{Fluorine Dimer}
\label{subsec:ch8:F2}
%%%%%%%%%%%%%%%%%%%%%%%%%%%%%%%%%%%%%%%%%%%%%%%%%%%%%%

% INTRODUCTION OF CHALLENGE
We turn next to the fluorine dimer as a molecular example with strong static and dynamic correlation effects across all geometries.
The nature of the electron correlation in \ce{F2} makes it a challenging test case for electronic structure methods,\cite{Laidig1987,Krylov2000,Kowalski2001} with the UHF approximation predicting an unbound potential.\cite{Laidig1987} 
In our previous work [Ref.~\onlinecite{Burton2019c}], we have reported that  NOCI wave functions built from the lowest HF states allow qualitatively correct bound potentials to be recovered with a similar accuracy to CASSCF calculations using a minimal active space.
However, like CASSCF, we found that NOCI predominantly captured static correlation effects and retained a significant error relative to the exact FCI energy.
The \ce{F2} binding curve therefore presents an essential test for assessing the performance of NOCI-PT2.
Given the additional electrons in \ce{F2} compared to \ce{H4}, the cc-pVDZ basis set is beyond the memory capabilities of our current NOCI-PT2 implementation and instead we use Pople's 6-31G basis set.\cite{Ditchfield1971}

% DESCRIPTION OF SCF, NOCI, and CAS
The SCF metadynamics approach was applied using an active space [see Ref.~\onlinecite{Burton2019c}] containing the valence occupied $\sigg$ and virtual $\sigu$ orbitals to locate eight HF states at a bond length of $\SI{4}{\angstrom}$.
Following relaxation in the full orbital space, these states mirror those found in Ref.~\onlinecite{Burton2019c}.
The lowest energy state is a doubly-degenerate symmetry-broken UHF solution representing two open-shell \ce{F} atoms, followed by the ``$\sigg^2$-like'' RHF ground state, the doubly-degenerate ``$\sigg \sigu$-like'' UHF states, a ``$\sigu^2$-like'' solution, and the doubly-degenerate symmetry-broken ``ionic'' RHF states.
These states are then followed across the full binding curve, and complex-valued h-HF states are identified beyond the Coulson--Fischer points where the symmetry-broken UHF and ionic RHF states coalesce with the $\sigg^2$-like and $\sigg^2$-like states respectively (see top right panel in Fig.~\ref{fig:f2_6-31g}).
Taking only the RHF ground state and two degenerate symmetry-broken (h-)UHF states, the simplest NOCI~(3) wave function recovers a qualitatively correct bound potential with a close correspondence to the equivalent CASSCF~(2,2) energy (see left panel in Fig.~\ref{fig:f2_6-31g}).
Here the CASSCF~(2,2) active space is defined using the same valence occupied $\sigg$ and virtual $\sigu$ orbitals as the active-space SCF metadynamics calculation.
While both NOCI~(3) and CASSCF~(2,2) vastly underestimate the depth of the potential well, adding the remaining (h-)HF states in the NOCI~(8) expansion significantly deepens the well, but leads to an overestimate of the dissociation energy.

% DESCRIPTION OF NOCI-PT2
Despite providing a qualitatively correct binding curve, NOCI~(3) and NOCI~(8)  fail to capture any significant  dynamic correlation effects and have an absolute error relative to FCI ranging from $150$ to $200~\mEh$.
Adding dynamic correlation using the NOCI-PT2 correction significantly improves both the NOCI~(3) and NOCI~(8) energy and reduces the error to within $9~\mEh$ of the FCI result across the full binding curve, providing a similar accuracy as the CASPT2~(2,2) energy (left panel in Fig.~\ref{fig:f2_6-31g}).
Furthermore, the NOCI-PT2 correction provides an improved estimate of the equilibrium bond length compared to the reference NOCI binding curve.
In contrast, the NOCI-MP2 energy using the same NOCI~(3) reference wave function closely matches NOCI-PT2~(3) in the equilibrium regime, but fails catastrophically when the MP2 corrections to the constituent HF states break down at larger bond lengths.\cite{Laidig1987}
We were unable to compute a NOCI-MP2 binding curve using the NOCI~(8) reference wave function as the perturbed NOCI-MP2 overlap matrix contained unphysical negative eigenvalues for all bond lengths.

%============================================================%
\begin{figure}[tb!]
	\includegraphics[width=\linewidth]{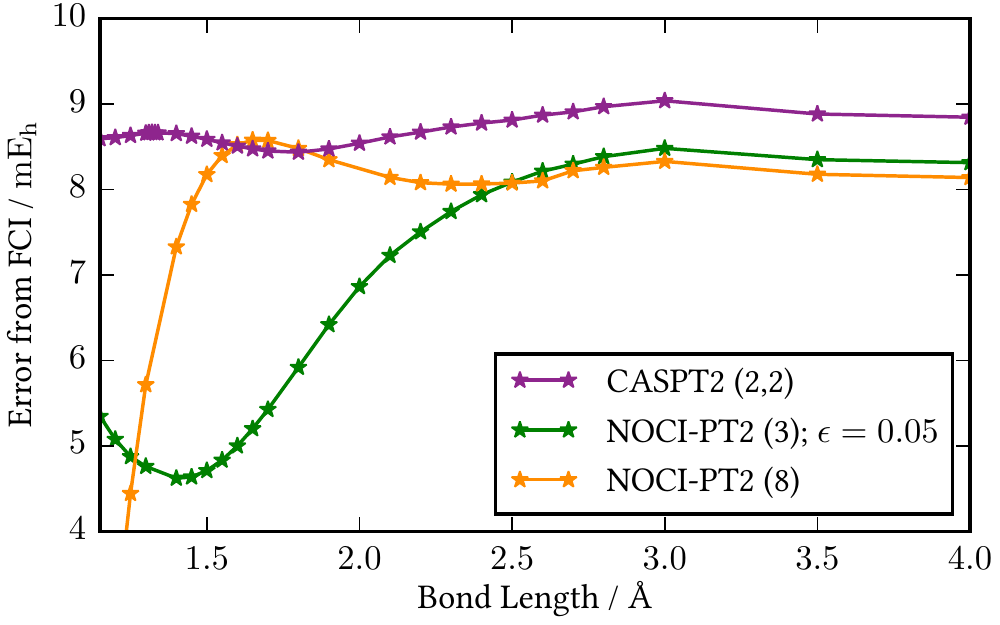}
	\caption[NOCI-PT2 binding curve for the fluorine dimer]
	{   \label{fig:f2_6-31g_npe}
		Comparison of the non-parallelity error for the NOCI-PT2 and CASPT2~(2,2) binding curves of \ce{F2} (6-31G) relative to FCI.
	}
\end{figure}
%============================================================%

% DISCUSSION OF IMAGINARY SHIFT AGAIN
The \ce{F2} binding curve is also not immune to the potential for intruder-state singularities, as demonstrated by the kink in the NOCI-PT2~(3) energy at $\SI{2.5}{\angstrom}$ in Fig.~\ref{fig:f2_6-31g}.
Since this kink occurs at a bond length where all the HF solutions are real, the corresponding singularity must arise from the relative energy and interaction between general
multiple HF solutions, rather than the inclusion of complex h-HF solutions.
Introducing an imaginary shift with $\epsilon = 0.05~\Eh$ allows this kink to be removed with no discernible distortion to the binding curve away from the singularity (bottom right panel in Fig.~\ref{fig:f2_6-31g}). 
Again, the shift required to remove the singularities in the NOCI-PT2 energy appears to be an order of magnitude smaller than those used in other multireference perturbation theories.\cite{Chang2012} 
Alternatively, the NOCI-PT2~(8) binding curve appears to be free from any intruder-state singularities.
This observation demonstrates that increasing the size of the NOCI expansion basis can avoid intruder states in a similar way as increasing the size of the CASSCF active space used in CASPT2 calculations.\cite{Pulay2011}

% SIZE-CONSISTENCY
Since the NOCI description is not exact at any geometry, we can now assess the size-consistency of the NOCI and NOCI-PT2 energies.
Allowing symmetry-breaking at the UHF level allows the HF energy to dissociate to the HF energy of two separated \ce{F} atoms.
Including these symmetry-broken UHF states in the NOCI wave functions then allows the NOCI energy to size-consistently dissociate to the same UHF energy for the separated atoms, as shown in Fig.~\ref{fig:f2_6-31g}.
In contrast, the CASSCF~(2,2) energy dissociates to the restricted open-shell HF (ROHF) energy of the separated atoms as the inactive orbitals are fixed to be doubly occupied. 
However, the NOCI-PT2 and CASPT2 binding curves do not dissociate to the corresponding unrestricted MP2 (UMP2) or restricted open-shell MP2 (ROMP2) energies respectively.
%Furthermore, building a NOCI~(2) wave function using only the two size-consistent symmetry-broken UHF states (not shown) also fails to dissociate to the UMP2 energy, suggesting that size-inconsistent terms are inherent in the NOCI-PT2 expansion. 
It is not entirely clear whether assessing the size-consistency of NOCI-PT2 using the UMP2 energy is a fair comparison, particularly as UMP2 itself fails to predict a qualitatively correct binding curve.\cite{Laidig1987}
Regardless, it is likely that size-inconsistent terms are inherent in the NOCI-PT2 expansion, and we intend to identify and remove these terms in future investigations. %, but is beyond the scope of the current paper.
For now, we note that the lack of exact size-consistency has not severely limited the applicability or popularity of CASPT2,\cite{Rintelman2005} and we believe that the same will be true for NOCI-PT2.

% PARALLELITY
While NOCI-PT2 does not quite recover the exact FCI energy, the relative shape of the NOCI-PT2 binding curves can be assessed using the non-parallelity error shown in Fig.~\ref{fig:f2_6-31g_npe}.
The NOCI-PT2~(3) energy shows a relatively high degree of non-parallelity, with the error relative to the FCI energy ranging from around $8~\mEh$ at large bond lengths to $5~\mEh$ in the equilibrium regime.
In contrast, the NOCI-PT2~(8) energy has a much more consistent error of around $8$--$9~\mEh$ across the majority of the binding curve, including the equilibrium bond length, although this error falls steeply when the nuclear repulsion starts to dominate the energy.
The variation in the NOCI-PT2~(8) error is comparable to the variation in the CASPT2~(2,2) error, and thus the two approaches produce similar relative binding curves.

%%%%%%%%%%%%%%%%%%%%%%%%%%%%%%%%%%%%%%%%%%%%%%%%%%%%%%%%%%%%%%%%%%%%%%%%%%%%%%%%%%%%%%%
\begin{squeezetable}
	\begin{table}[h!]
		\begin{ruledtabular}
			\begin{tabular}{ld{4.5}d{4.5} d{2.2}d{3.1}}
				&	\head{$E_{\text{min}} / \Eh$}	&	\head{$E (\SI{100}{\angstrom}) / \Eh$} 	& \head{$D_{\text{e}} / \mEh$}	& \head{\%$D_{\text{e}}(\text{FCI})$} 
				\Tstrut\\
				\hline 
				CASSCF~(2,2)			&	-198.73907	&	-198.72044		&	18.63		&	50.5
				\Tstrut\\
				NOCI~(3)					&  -198.73848	&   -198.72172       &   16.76		&    45.4
				\\
				NOCI~(8)        			&  -198.77411	&   -198.72175       &   52.36		&    141.8
				\\              
				CASPT2~(2,2)            & 	-198.92321	&  -198.88608		&	37.14		&   100.6
				\\
				NOCI-PT2~(3)			&	-198.92696	& -198.88660		&  40.36		& 109.3
				\\
				NOCI-PT2~(8)    		&  -198.92336 	& -198.88679		&  36.57		& 99.1
				\\ \hline\Tstrut
				FCI 						  &  -198.93176  	& -198.89485		& 36.91			& 100.0
			\end{tabular}
		\end{ruledtabular}
		\caption{Comparison of the dissociation energy of \ce{F2} (6-31G) using CASPT2 and NOCI-PT2 relative to the exact FCI result. 
			The energy $E_{\text{min}}$ is taken as the minimum of the points sampled in Fig.~\ref{fig:f2_6-31g}.}
		\label{tab:F2_6-31G_De}
	\end{table}
\end{squeezetable}
%%%%%%%%%%%%%%%%%%%%%%%%%%%%%%%%%%%%%%%%%%%%%%%%%%%%%%%%%%%%%%%%%%%%%%%%%%%%%%%%%%%%%%%\\

%============================================================%
\begin{figure*}[tb!]
	\includegraphics[width=\linewidth]{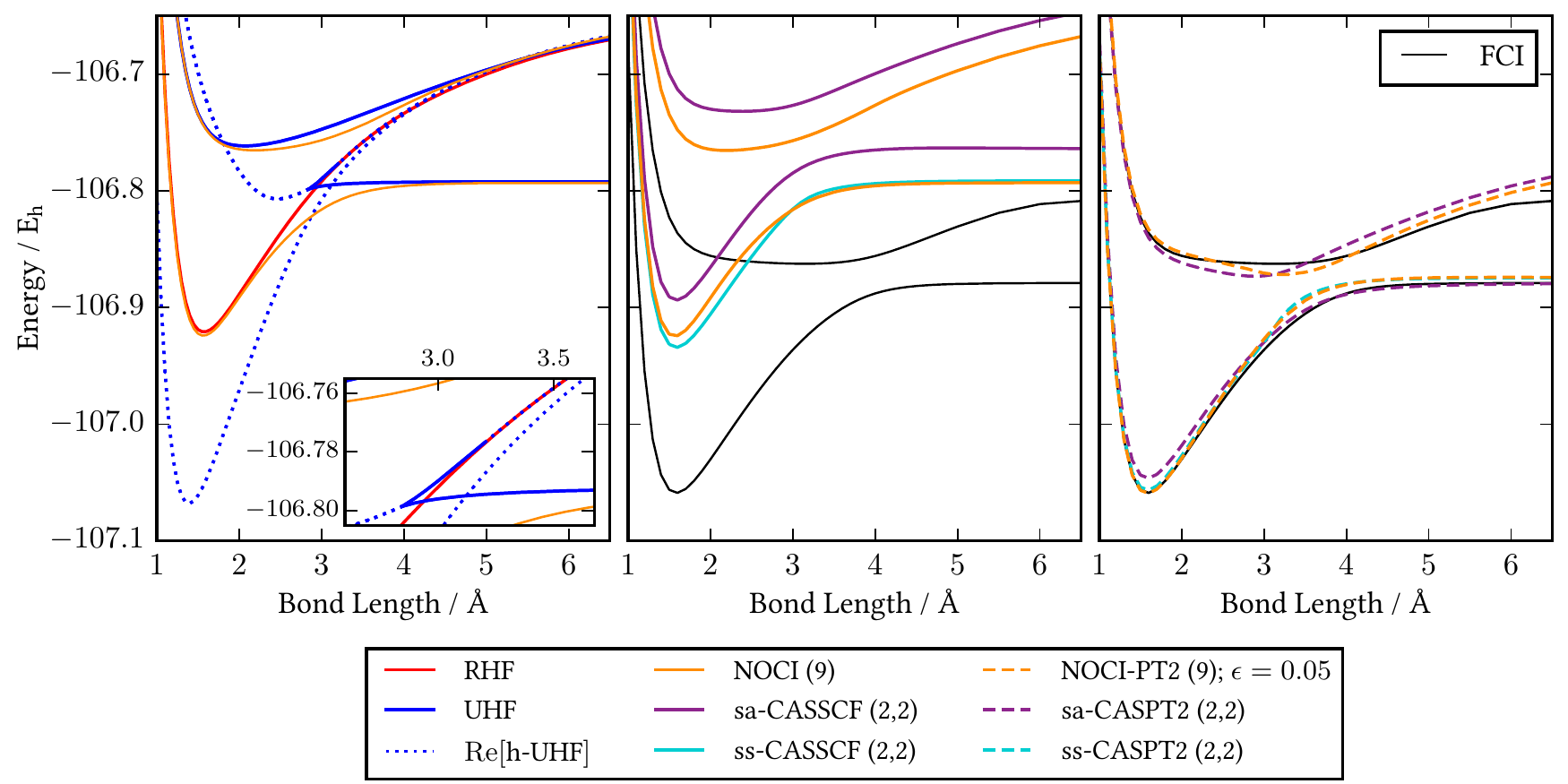}
	\caption
	{   \label{fig:lif_6-31g}
		\textit{Left:} Nine (h-)HF states identified for the \ce{LiF} binding curve include three doubly-degenerate UHF states, the ionic RHF configuration, and a doubly-degenerate h-UHF state with complex coefficients across all bond lengths. 
		The h-HF energy may be complex-valued in general and thus there is no condition for the real component of the h-UHF energy to be variational (see Ref.~\onlinecite{Burton2018}).
		\textit{Middle:} NOCI~(9) ground and excited states computed using the nine (h-)HF solutions compared to sa-CASSCF~(2,2) and the ss-CASSCF~(2,2) ground state.
		\textit{Right:}	The imaginary-shifted NOCI-PT2~(2) correction with $\epsilon=0.5~\Eh$ recovers most of the remaining dynamic correlation, improving the position and shape of the avoided crossing.
	}
\end{figure*}
%============================================================%

% DISSOCIATION ENERGY
Encouraged by the non-parallelity errors for NOCI-PT2, we compare the dissociation energy computed using the minimum energy of the points sampled in Fig.~\ref{fig:f2_6-31g} relative to the energy at a bond-length of $\SI{100}{\angstrom}$, as shown in Table~\ref{tab:F2_6-31G_De}.
The reference NOCI~(3) and NOCI~(8) wave functions either underestimate or overestimate the dissociation energy by almost $50~\%$ respectively.
Adding the NOCI-PT2 correction significantly improves the dissociation energy, with the larger NOCI-PT2~(8) expansion space reaching chemical accuracy within $0.5~\mEh$ of the FCI result.
Furthermore, while NOCI-PT2~(8) and CASPT2~(2,2) predict the dissociation energy with a similar level of accuracy, the NOCI-PT2~(8) approach has the advantage of providing marginally improved absolute energies along the full binding curve.
These results reinforce the quantitative accuracy gained by the NOCI-PT2 correction relative to the NOCI reference wave function, and further illustrate the systematic improvement provided by increasing the size of the NOCI expansion space.

%%%%%%%%%%%%%%%%%%%%%%%%%%%%%%%%%%%%%%%%%%%%%%%%%%%%%%
\subsection{Lithium Fluoride}
\label{subsec:LithiumFluoride}
%%%%%%%%%%%%%%%%%%%%%%%%%%%%%%%%%%%%%%%%%%%%%%%%%%%%%%

% GENERAL INTEREST
Our final example, the \ce{LiF} molecule, has long been used as a testing ground for multireference methods.\cite{Bauschlicher1988,Werner1981,Malrieu1995,Serrano-Andres2005,Sivalingam2016,Mayhall2014,Thom2009}
The \ce{LiF} ground state is known to be ionic in character for short bond lengths, but the molecule covalently dissociates to neutral open-shell \ce{Li} and \ce{F} atoms.\cite{Bauschlicher1988}
Since both configurations share the same $^{\small{1}}\Sigma^{+}$ symmetry, the transition from predominantly ionic character to neutral character is marked by a sharp avoided crossing.
Like all single-bond breaking processes, predicting the full \ce{LiF} binding curve requires a multireference approach that can account for static correlation in the  dissociation limit.
However, predicting the exact location of the avoided crossing also relies on an adequate description of dynamic correlation that preferentially stabilises the ionic configuration.

% ORBITAL RELAXATION AND NOCI
The ability to include orbital relaxation effects and diabatic states in NOCI makes it a promising method for representing the ionic and neutral configurations in \ce{LiF}.\cite{Thom2009,Mayhall2014,Nite2019b}
Indeed, diabatic HF states corresponding to the ionic and neutral configurations have previously been identified by Thom and Head-Gordon in Ref.~\onlinecite{Thom2009} and were found to  reproduce an adiabatic avoided crossing when combined using NOCI.
However, these NOCI expansions could not be applied across the full binding curve as the covalent HF state was found to vanish at a Coulson--Fischer point.\cite{Thom2009}
We note that Nite and Jim\'{e}nez-Hoyos later identified and used the same HF states in their NOCISD calculations on \ce{LiF},\cite{Nite2019b} and we speculate that they were also limited by the disappearance of the covalent solution.
Alternatively, the spin-flip NOCI approach avoided issues associated with disappearing solutions by using partially optimised determinants obtained through the spin-flip philosophy, leading to smooth adiabatic states across all geometries.\cite{Mayhall2014} 

% SCF DISCUSSION
The development of h-HF theory\cite{Hiscock2014,Burton2016,Burton2018} now allows the neutral HF state to be extended across all molecular structures.
Using the 6-31G basis set,\cite{Ditchfield1971} we identify nine significant (h-)HF solutions including those described by Thom and Head-Gordon,\cite{Thom2009} as shown in Fig.~\ref{fig:lif_6-31g} (left panel).
The HF ground state at the equilibrium geometry is a RHF solution representing the ionic bonding \ce{Li^{+}\bond{-}F^{-}} configuration.
In contrast, the ground state in the dissociation limit corresponds to a doubly-degenerate symmetry-broken UHF solution representing to the covalent \ce{Li^{$\upharpoonright$}\bond{-}F^{$\downharpoonright$}} and \ce{Li^{$\downharpoonright$}\bond{-}F^{$\upharpoonright$}} configurations.
These diabatic RHF and UHF solutions cross at around $\SI{3}{\angstrom}$, but the neutral states then coalesce with a third UHF state and vanish for shorter bond lengths (see inset in the left panel of Fig.~\ref{fig:lif_6-31g}). 
Surprisingly, the third UHF state only exists for a short range of bond lengths before first coalescing with the ionic RHF solution, then extending as a complex h-UHF solution, and finally coalescing with another h-UHF state that is complex-valued across all geometries.\footnote{Holomorphic HF states that are complex across all geometries have previously been described as ``dormant'' states in Ref.~\protect\onlinecite{Burton2018}.}
The canonical orbitals for this third UHF state differ from the covalent and ionic configurations by only one electron that occupies a delocalised bonding orbital between the \ce{Li} and \ce{F} atoms.
This state therefore represents a partial electron transfer that mirrors the HF solutions previously identified in the \ce{C7H6F4^{+}} electron transfer model.\cite{Jensen2018}
Finally, the highest-energy doubly-degenerate UHF state that we identified contains a similar configuration, although the delocalised electron now occupies an anti-bonding orbital.
A comprehensive discussion on the coalescence points and h-HF energies of these stationary states is available in the Supporting Information.

% NOCI
Taking a linear combination of these nine (h-)HF states yields adiabatic NOCI~(9) states that recover an avoided crossing, as shown in the middle panel of Fig.~\ref{fig:lif_6-31g}.
The $1\,^{1}\Sigma^{+}$ state smoothly interpolates between the ionic ground state at short bond lengths and the covalent ground state in the dissociation limit.
In contrast, recovering these adiabatic states using CASSCF~(2,2) requires a state-averaged CASSCF (sa-CASSCF) formalism.\cite{Werner1981}
Using an equal weighting for each state, we find that sa-CASSCF~(2,2) gives a significant energy penalty relative to the NOCI~(9) ground and excited state.
The effect of state-averaging can be highlighted by comparing to the state-specific CASSCF (ss-CASSCF) ground state, and indeed we find that the ss-CASSCF~(2,2) energy shows a much closer correspondence to the NOCI~(9) energy.
While both NOCI~(9) and sa-CASSCF~(2,2) recover adiabatic states, the corresponding avoided crossing is not sharp enough and occurs at shorter bond lengths than the exact FCI result.
Furthermore, the ground $1\,^{1}\Sigma^{+}$ state and excited $2\,^{1}\Sigma^{+}$ state computed using NOCI~(9) and sa-CASSCF~(2,2) retain a significant absolute error relative to their FCI counterparts.

%============================================================%
\begin{figure}[tb!]
	\includegraphics[width=\linewidth]{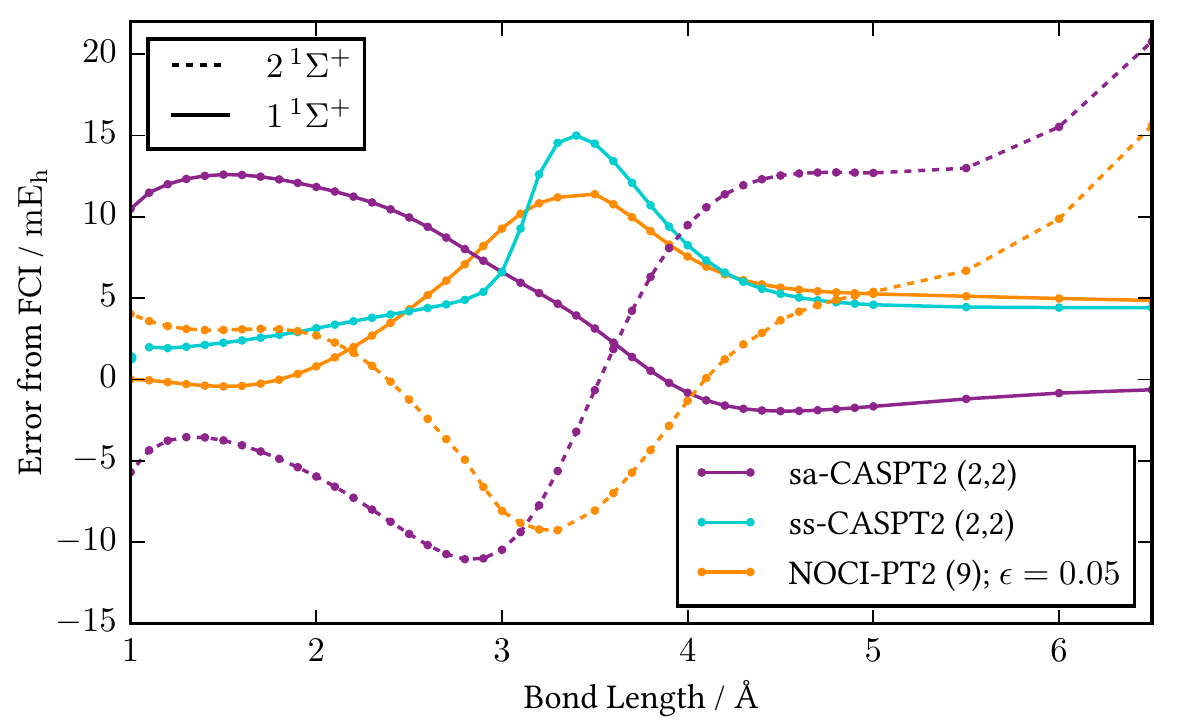}
	\caption
	{   \label{fig:lif_6-31g_npe}
		Comparison of the non-parallelity error for the  NOCI-PT2~(9)  and CASPT2~(2,2) ground and excited $^{1}\Sigma^{+}$ states of \ce{LiF} (6-31G) relative to the exact FCI energy.
		The ss-CASSCF calculations appeared to experience convergence issues associated with multiple solutions at very short bond lengths.
	    As a result, we were unable to follow the desired solution between $\SI{1.1}{\angstrom}$ and $\SI{1.0}{\angstrom}$.
	}
\end{figure}
%============================================================%

% NOCI-PT2
Adding dynamic correlation through NOCI-PT2 significantly lowers the energy of both the ground and excited state, as shown in Fig.~\ref{fig:lif_6-31g_npe}.
Using an imaginary shift of $\epsilon=0.05~\Eh$ to avoid any singularities, NOCI-PT2 reduces the error to within $15~\mEh$ of the exact FCI energies across the full binding curve.
NOCI-PT2 and ss-CASPT2 predict remarkably similar ground-state  binding curves, although NOCI-PT2 provides greater accuracy in the equilibrium region.
Furthermore, NOCI-PT2 provides a significantly improved estimate for the position and sharpness of the avoided crossing.
In contrast, the sa-CASPT2 ground- and excited-state energies show a high degree of non-parallelity and provide a worse estimate of the avoided crossing, presumably due to the inferior quality of the reference wave function for both states.
These results demonstrate one major advantage of the NOCI-PT2 approach: NOCI-PT2 provides ground- and excited-state energies with equivalent accuracy to a state-specific CASPT2 calculation while avoiding the need for any arbitrary state-averaging. 

%%%%%%%%%%%%%%%%%%%%%%%%%%%%%%%%%%%%%%%%%%%%%%%%%%%%%%
\section{Concluding Remarks}
\label{sec:ConcludingRemarks}
%%%%%%%%%%%%%%%%%%%%%%%%%%%%%%%%%%%%%%%%%%%%%%%%%%%%%%

% DERIVING THE THEORY
We have introduced a rigorous second-order diagonalise-then-perturb correction that allows dynamic correlation to be captured on top of the statically correlated NOCI wave function.
Our NOCI-PT2 theory uses a generalised Fock reference Hamiltonian and forms the rigorous nonorthogonal extension to conventional multireference perturbation theory.
Despite the lack of any well-defined Hilbert space partitioning for the NOCI reference wave function, we have shown that the first-order interacting space can be constructed from the combined set of single and double excitations built from each reference determinant in the NOCI expansion.
This first-order interacting space allows an approximate second-order energy correction to be defined in terms of a polynomially-scaling number of perturbation determinants. 

% RESULTS SO FAR
By considering a series of challenging molecular systems exhibiting both static and dynamic correlation, we have demonstrated that NOCI-PT2 provides similar quantitative accuracy to an equivalent CASPT2 calculation.
In particular, we have found that NOCI-PT2 recovers chemical accuracy for the dissociation energy of \ce{F2} and significantly improves the shape of the avoided crossing in \ce{LiF} relative to the reference NOCI calculation.
Furthermore, NOCI-PT2 consistently predicts more accurate energy surfaces than the previous NOCI-MP2 approach,\cite{Yost2013,Yost2016,Yost2019} while the latter can fail completely if negative eigenvalues appear in the NOCI-MP2 overlap matrix.
NOCI-PT2 therefore provides a reliable multireference approach for predicting quantitative energies while exploiting the numerous advantages of the NOCI framework.

% IMAGINARY SHIFT
To remove potential intruder-state singularities in the NOCI-PT2 energy, we have introduced an imaginary-shifted variant of the NOCI-PT2 equations.
Our imaginary-shifted approach employs the explicit complex-valued reference Hamiltonian $\Href \rightarrow \Href + \I \epsilon \cQ$ instead of the approximate forms used to retain real arithmetic elsewhere.\cite{Forsberg1997,Tsuchimochi2019}
We find that this complex implementation requires shift values that are an order of magnitude smaller than those used in approximate real-valued approaches.\cite{Chang2012}
As a result, the explicit imaginary shift can recover smooth potential energy surfaces with far less distortion away from intruder-state singularities.
Given the ease of implementing complex arithmetic in modern computer programming, we suggest revisiting the use of a complex imaginary-shift equation in orthogonal multireference perturbation theory.

% EXCITED STATES
Among the most promising aspects of our NOCI-PT2 results is the accuracy of the NOCI-PT2 ground and excited $^1\Sigma^{+}$ states in the avoided crossing of \ce{LiF}.
We have found that NOCI-PT2 produces a similar energy to a state-specific CASPT2 ground state while also providing the same accuracy for the excited state.
In contrast, computing both the ground and excited states using CASPT2 requires a state-averaged formalism that reduces the quality of both energy surfaces and relies on the arbitrary definition of optimisation weights for each state.
Our results therefore indicate that NOCI-PT2 can provide accurate and efficient state-specific multireference excited-state energies without relying on any state-averaging procedures.
Furthermore, the adiabatic NOCI-PT2 excited states can provide chemical interpretation of avoided crossings through the use of diabatic HF states in the NOCI expansion.
We will continue to investigate the wider application of NOCI-PT2 to excited-state energies in future publications.

% COMPUTATIONAL DIRECTIONS
%At present, our computational NOCI-PT2 approach is limited by two main factors.
%Firstly, explicitly building and storing the full NOCI-PT2 matrices creates a memory bottleneck, while the cost of the generalised Slater--Condon rules\cite{MayerBook} prevents matrix elements from being computed on the fly.
%To extend NOCI-PT2 to larger systems using an iterative on-the-fly approach, we intend to compute matrix elements through the more efficient Wick's theorem, where each element is expressed in terms of the reference determinants.\cite{RingBook}
%Secondly, the presence of redundancies in the first-order interacting space can lead to slow iterative convergence of the NOCI-PT2 linear problem.
%Introducing a contraction scheme for the perturbing determinants may allow the effect of these redundancies to be minimised.
%A contracted NOCI-PT2 approach would also improve the computational scaling of NOCI-PT2 by reducing the dimensionality of the linear problem, and research in this direction is ongoing.

% CHEMICAL EXAMPLES
In summary, our nonorthogonal second-order multireference perturbation theory provides a rigorous and reliable approach for reaching quantitative energies using NOCI wave functions.
These developments establish the combined NOCI~/~NOCI-PT2 approach as a competitive alternative to the CASSCF~/~CASPT2 route for treating both static and dynamic correlation.
Ultimately, NOCI-PT2 now allows us to take full advantage of the NOCI framework and use multiple HF states to quantitatively understand molecular processes. % such as bond dissociation\cite{Thom2009,Burton2019c} or electron transfer.\cite{Jensen2018}

% ULTIMATE SUMMARY
%Ultimately, the rigorous NOCI-PT2 correction developed in this paper now establishes NOCI-based methods as a fully fledged alternative to conventional multireference electronic structure theory.

%%%%%%%%%%%%%%%%%%%%%%%%%%%%%%%%%%%%%%%%%%%%%%%%%%%%%%%%%%%%%%
\section*{Acknowledgements}
%%%%%%%%%%%%%%%%%%%%%%%%%%%%%%%%%%%%%%%%%%%%%%%%%%%%%%%%%%%%%%
H.G.A.B. thanks the Cambridge Trust for a Vice Chancellor's Award studentship and A.J.W.T.~thanks the Royal Society for a University Research Fellowship (UF110161). 
We also thank Charles Scott, Sandeep Sharma, Anthony Scemama and Pierre-Fran\c{c}ois Loos for useful discussions at various stages throughout the development of this work.

This work was performed using resources provided by the Cambridge Service for Data Driven Discovery (CSD3) operated by the University of Cambridge Research Computing Service (\href{www.csd3.cam.ac.uk}{www.csd3.cam.ac.uk}), provided by Dell EMC and Intel using Tier-2 funding from the Engineering and Physical Sciences Research Council (capital grant EP/P020259/1), and DiRAC funding from the Science and Technology Facilities Council (\href{www.dirac.ac.uk}{www.dirac.ac.uk}).

%%%%%%%%%%%%%%%%%%%%%%%%%%%%%%%%%%%%%%%%%%%%%%%%%%%%%%%%%%%%%%
\section*{Supporting Information}
%%%%%%%%%%%%%%%%%%%%%%%%%%%%%%%%%%%%%%%%%%%%%%%%%%%%%%%%%%%%%%
Analysis of the h-HF states for \ce{LiF} (6-31G), including a discussion on the coalescence behaviour and the h-HF energy of each state.

%%%%%%%%%%%%%%%%%%%%%%%%%%%%%%%%%%%%%%%%%%%%%%%%%%%%%%%%%%%%%%
\appendix
%%%%%%%%%%%%%%%%%%%%%%%%%%%%%%%%%%%%%%%%%%%%%%%%%%%%%%%%%%%%%%
\section{Derivation of Imaginary Shift Approximations}
\label{apdx:ImagShift}
%%%%%%%%%%%%%%%%%%%%%%%%%%%%%%%%%%%%%%%%%%%%%%%%%%%%%%%%%%%%%%

In this Appendix, we discuss the relationship between the rigorous imaginary shift Eq.~\eqref{eq:imagShiftH} and the approximate forms proposed in CASPT2\cite{Forsberg1997} and SUPT2.\cite{Tsuchimochi2019}
We demonstrate that the approximate forms used in CASPT2 and SUPT2 require diagonal approximations in the context of NOCI-PT2 that lead to extremely poor iterative convergence.

Following Section~\ref{subsec:ch8:ExternalSpace}, the first-order interacting space is constructed from the projected single and double excitations $\{\ket{\Omega^{wI}}\}$ defined in Eq.~\eqref{eq:FOIS}.
These expansion functions generally form a nonorthogonal contravariant basis [see Ref.~\onlinecite{HeadGordon1998}] and the corresponding contravariant metric tensor  is defined through the overlap matrix \eqref{eq:NOCIPT2_smat} as
\begin{equation}
	Q^{wI, xJ} = 	\brkt{\Omega^{wI}}{\Omega^{xJ}}.
	\label{eq:MetTens}
\end{equation}
For the sake of brevity, from here-on we replace the indices ($wI,xJ,\dots$) with the compound indices ($p,q,r,s,\dots$).
Furthermore, to allow easy comparison with previous expressions in CASPT2\cite{Forsberg1997} and SUPT2,\cite{Tsuchimochi2019} we assume real-valued NOCI-PT2 matrix elements and first-order wave function coefficients,  although the results extend to the general case where these may become complex.
Finally, we employ the implicit summation convention for any symbol that appears as both a covariant and contravariant index.\cite{HeadGordon1998}

The rigorous imaginary-shifted linear problem \eqref{eq:imShiftLinProb} is expressed in terms of the first-order interacting space as
\begin{equation}
	\qty( M^{pq} + \I \epsilon\, Q^{pq} ) a_q = - V^{p}.
	\label{eq:apdx:imShiftLinProb}
\end{equation}
Decomposing the first-order wave function coefficients into real and imaginary components as $a_q = x_q + \I y_q$ leads to the simultaneous equations
\begin{subequations}
\begin{align}
	M^{pq} x_q - \epsilon\, Q^{pq}  y_q &= - V^{p},
	\label{eq:apdx:imShiftLinProbR}
	\\
	M^{pq} y_q + \epsilon\, Q^{pq} x_q &= 0.
	\label{eq:apdx:imShiftLinProbZ}
\end{align}
\end{subequations}
Note that for complex-valued matrix elements, the ``real'' component $x_q$ is itself complex-valued and represents the approximate solution to the unshifted NOCI-PT2 equations, while the ``imaginary'' component $y_q$ represents the distortion induced by the imaginary shift.
In CASPT2 and SUPT2, only the real components $x_q$ are used to evaluate the imaginary-shifted second-order energy.\cite{Forsberg1997,Tsuchimochi2019}

Using the inverse relationship
\begin{equation}
[M^{-1}]_{rp} M^{pq} = \delta_{r\cdot}^{\cdot q}.
\end{equation}
allows Eq.~\eqref{eq:apdx:imShiftLinProbZ} to be explicitly solved to give
\begin{equation}
	y_r = - \epsilon [M^{-1}]_{rp} Q^{pq} x_q.
\end{equation}
The imaginary components $y_q$ can then be eliminated from Eq.~\eqref{eq:apdx:imShiftLinProbR} to give a linear problem in terms of only the real component of the first-order wave function coefficients as
\begin{equation}
	\qty(M^{pq} + \epsilon^2\, [M^{-1}]^{pq}) x_q = - V^{p},
	\label{eq:Modified1}
\end{equation}
where we have used the properties of the metric tensor~\eqref{eq:MetTens} to obtain
\begin{equation}
	[M^{-1}]^{pq} = Q^{pr} [M^{-1}]_{rs} Q^{sq}.
\end{equation}

In principle, solving Eq.~\eqref{eq:Modified1} allows the components $x_q$ to be identified using only real arithmetic. 
However, the presence of $\bM^{-1}$ generally makes this approach untenable.
Instead, taking the diagonal approximation 
\begin{equation}
	[M^{-1}]^{pq} \approx \frac{\delta^{pq}}{M^{pp}}
\end{equation}
leads to the route proposed in the original imaginary-shifted CASPT2 approach.\cite{Forsberg1997}
This diagonal approximation will only perform well if  $\bM$ is a sparse and diagonally dominant matrix, which will generally not be the case for the NOCI-PT2 first-order interacting determinants.
Our preliminary NOCI-PT2 investigations have confirmed that using this diagonal approximation leads to extremely poor iterative convergence of Eq.~\eqref{eq:Modified1}.

Alternatively, one may attempt to remove the dependence on $\bM^{-1}$ by pre-multiplying Eq.~\eqref{eq:Modified1} with $M^{rs} Q_{sp}$ to give
\begin{equation}
	\qty(M^{rs} Q_{sp} M^{pq} + \epsilon^2\, Q^{rq}) x_q = - M^{rs} Q_{sp} V^{p},
	\label{eq:Modified2}
\end{equation}
where we have exploited the relationship
\begin{equation}
	M^{rs} Q_{sp} [M^{-1}]^{pq} = Q^{r q}.
\end{equation}
The quadratic form \eqref{eq:Modified2} is the nonorthogonal extension of the imaginary-shifted approach introduced for the SUPT2.\cite{Tsuchimochi2019}
Note that the use of a ``projection-after-excitation'' formalism in SUPT2 means that the underlying basis is orthogonal and $Q^{pq} = Q_{pq} = \delta_{pq}$.
In contrast, the presence of the inverse overlap matrix elements $Q_{sp}$ for the nonorthogonal functions \eqref{eq:FOIS} makes solving Eq.~\eqref{eq:Modified2} for NOCI-PT2 as difficult as solving Eq.~\eqref{eq:Modified1}.
Furthermore, introducing the quadratic dependence on $\bM$ will amplify any null-space effects arising from linear redundancies in the first-order interacting space.
While we may approximately solve Eq.~\eqref{eq:Modified2} using the diagonal approximations $Q_{pq} \approx \delta_{pq}$ and $Q^{pq} \approx \delta^{pq}$, our preliminary investigations have confirmed that this also leads to extremely poor iterative convergence of the NOCI-PT2 linear problem.

In summary, there are no major computational advantages in using either of the previous imaginary-shift approximations defined in CASPT2\cite{Forsberg1997} or SUPT2.\cite{Tsuchimochi2019}
Since generally the NOCI-PT2 equations already require complex arithmetic, it is more robust and efficient to use the explicit imaginary-shifted equations given by Eqs.~\eqref{eq:imShiftLinProb} and \eqref{eq:imShift}.

%%%%%%%%%%%%%%%%%%%%%%%%%%%%%%%%%%%%%%%%%%%%%%%%%%%%%%%%%%%%%%
\section*{References}
\bibliography{NOCI_PT2}
%%%%%%%%%%%%%%%%%%%%%%%%%%%%%%%%%%%%%%%%%%%%%%%%%%%%%%%%%%%%%%

\end{document}